\documentclass[10pt,conference]{IEEEtran}
\usepackage{cite}
\usepackage{amsmath,amssymb,amsfonts}
\usepackage{algorithmic}
\usepackage{graphicx}
\usepackage{textcomp}
\usepackage{xcolor}
\usepackage[hyphens]{url}
\usepackage{fancyhdr}
\usepackage{hyperref}
\usepackage[capitalise,noabbrev,nameinlink]{cleveref}
\usepackage{comment}
\usepackage{tcolorbox}
\usepackage{multirow}
\usepackage{soul}

\IEEEoverridecommandlockouts

\newcommand{\PRESS}{PROTEAS}
\newcommand{\TODO}[1]{\textcolor{red}{TODO:#1}}
\newcommand{\redx}[1]{\textcolor{red}{#1}}
\newcommand{\ignore}[1]{}
\newcommand{\revision}[1]{\textcolor{red}}

\title{\vspace{-0.15in}Probabilistic Tracker Management Policies for Low-Cost and Scalable Rowhammer Mitigation}

\author{%
\begin{tabular}{c} Aamer Jaleel \\ NVIDIA \\ ajaleel@nvidia.com \end{tabular} \and
\begin{tabular}{c} Stephen W. Keckler \\ NVIDIA \\ skeckler@nvidia.com \end{tabular} \and
\begin{tabular}{c} Gururaj Saileshwar\textsuperscript{$*$}\thanks{\textsuperscript{$*$}Gururaj contributed to this work while he was affiliated with NVIDIA.} \\ University of Toronto \\ gururaj@cs.toronto.edu \end{tabular} }
\date{}

\usepackage{tikz}
\newcommand*\circled[1]{\tikz[baseline=(char.base)]{
            \node[shape=circle,draw,inner sep=2pt,fill=black, text=white] (char) {#1};}}

\fancypagestyle{camerareadyfirstpage}{%
  \fancyhead{}
  
  \fancyhead[C]{
    \ifdefined\aeopen
    \parbox[][12mm][t]{13.5cm}{}%\hpcayear{} IEEE International Symposium on High-Performance Computer Architecture (HPCA)}    
    \else
      \ifdefined\aereviewed
      \parbox[][12mm][t]{13.5cm}{}%\hpcayear{} IEEE International Symposium on High-Performance Computer Architecture (HPCA)}
      \else
      \ifdefined\aereproduced
      \parbox[][12mm][t]{13.5cm}{}%\hpcayear{} IEEE International Symposium on High-Performance Computer Architecture (HPCA)}
      \else
      \parbox[][0mm][t]{13.5cm}{}%\hpcayear{} IEEE International Symposium on High-Performance Computer Architecture (HPCA)}
    \fi 
    \fi 
    \fi 
    \ifdefined\aeopen 
      \includegraphics[width=12mm,height=12mm]{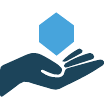}
    \fi 
    \ifdefined\aereviewed
      \includegraphics[width=12mm,height=12mm]{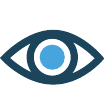}
    \fi 
    \ifdefined\aereproduced
      \includegraphics[width=12mm,height=12mm]{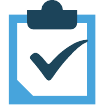}
    \fi
  }
  \fancyfoot[C]{}
}
\begin{document}
\maketitle

%Enables the camera ready header and footer
%\ifdefined\hpcacameraready 
%  \thispagestyle{camerareadyfirstpage}
%  \pagestyle{empty}
%\else
%  \thispagestyle{plain}
%  \pagestyle{plain}
%\fi
\thispagestyle{plain}
\pagestyle{plain}
\pagenumbering{arabic}

\newcommand{\hpcaheight}{0mm}
\ifdefined\eaopen
\renewcommand{\hpcaheight}{12mm}
\fi

\newcommand{\aj}[1]{{\footnotesize\color{red}[AJ: #1]}}
  
%%%%%%%%%%%%%%%%%%%%%%%%%%%%%%%%%%%%%%%%
%%%%%%%% -- PAPER CONTENT STARTS -- %%%%%%%%%

\begin{abstract}
This paper focuses on mitigating DRAM Rowhammer attacks.  In recent years, solutions like TRR have been deployed in DDR4 DRAM to track aggressor rows and then issue a mitigative action by refreshing neighboring victim rows. Unfortunately, such in-DRAM solutions are resource-constrained (only able to provision few tens of counters to track aggressor rows) and are prone to thrashing based attacks, that have been used to fool them. Secure alternatives for in-DRAM trackers require tens of thousands of counters. 

In this work, we demonstrate secure and scalable rowhammer mitigation using resource-constrained trackers. Our key idea is to manage such trackers with probabilistic management policies (PROTEAS). PROTEAS includes component policies like request-stream sampling and random evictions which enable thrash-resistance for resource-constrained trackers. We show that PROTEAS can secure small in-DRAM trackers (with 16 counters per DRAM bank) even when Rowhammer thresholds drop to 500 while incurring less than 3\% slowdown. Moreover, we show that PROTEAS significantly outperforms a recent similar probabilistic proposal from Samsung (called DSAC) while achieving 11X - 19X the resilience against Rowhammer.
\end{abstract}
%\vspace{-0.5in}
\section{Introduction}
DRAM scaling has led to higher DRAM capacities by packing DRAM cells more closely. This has increased inter-cell interference, leading to the problem of Rowhammer~\cite{kim2014flipping}, whereby rapid activations of one DRAM row can cause charge leakage and bit-flips in neighboring rows. Such Rowhammer bit-flips are not just a reliability problem, but also a major security threat. 
Numerous studies have illustrated exploits using Rowhammer~\cite{seaborn2015exploiting, frigo2020trrespass, gruss2018another, aweke2016anvil, cojocar2019eccploit, gruss2016rhjs, vanderveen2016drammer, kwong2020rambleed}.
Moreover, the vulnerability is worsening with each DRAM generation. 
The number of activations required to induce bit-flips, called the Rowhammer threshold (TRH), has dropped from 140K in DDR3~\cite{kim2014flipping} to just 4.9K in LPDDR4~\cite{kim2020revisitingRH} over the last decade. Thus, there is a growing need for effective and scalable mitigations, as TRH is expected to drop further. 

Mitigating Rowhammer effects within the DRAM module has been an ongoing challenge.
Since the release of DDR4 in 2015, DRAM manufacturers have deployed an in-DRAM mitigation called Targeted Row Refresh (TRR). 
TRR and many subsequent solutions rely on a \textit{tracking} mechanism to identify rapidly activated rows or aggressor rows and then issue a \textit{mitigative action} by refreshing the neighboring victim rows~\cite{hassan2021UTRR}.
The tracker typically consists of a group of counters within each DRAM bank that counts row activations and issues mitigations in the background of regular refresh commands when they are issued to DRAM by the memory controller.

\begin{figure*}[!htb]
    \centering
    \vspace{-0.1 in}
    \includegraphics[width=6.5in]{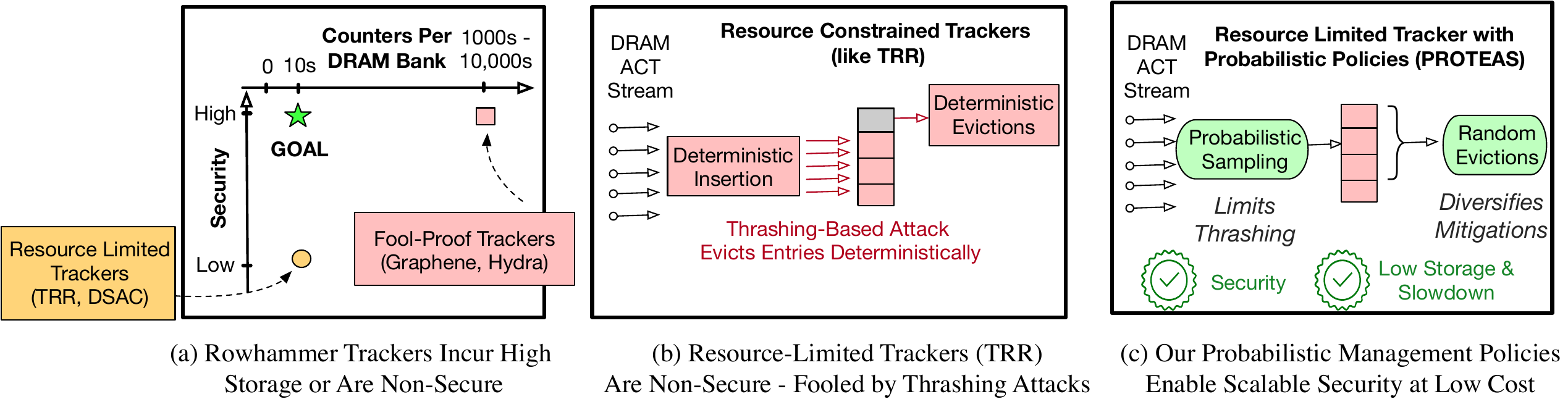}
    \vspace{-0.1 in}
    \caption{(a) As Rowhammer thresholds drop below 1K, Rowhammer trackers \ignore{that provide deterministic security guarantees }require 1000s to 10,000s of counters to be secure, while resource-constrained trackers (like TRR or DSAC) with tens of counters, which are practical within the logic area in DRAM, provide little to no security. (b) This is because trackers with tens of counters are vulnerable to thrashing-based attacks, which can easily evict tracked entries to fool the tracker, exploiting its deterministic management. (c) \PRESS{} provides strong security with a small 16-entry tracker, via thrash resistance through probabilistic sampling and random evictions.
    }
    \label{fig:intro}
    \vspace{-0.2 in}
\end{figure*}

Unfortunately, given the limited storage capabilities within the logic space of a DRAM module, TRR implementations often store less than 32 counters per DRAM bank~\cite{hassan2021uncovering,jattke2021blacksmith,de2021smash}. 
Such limited capacity of in-DRAM trackers has made them significantly vulnerable to thrashing-based attacks such as TRRespass~\cite{frigo2020trrespass} and Blacksmith~\cite{jattke2021blacksmith}, which ensure aggressor rows are evicted from the tracker by activating a larger number of rows than the tracker capacity, as shown in \cref{fig:intro}(b). Such thrashing-based attacks can continue to activate untracked rows far beyond TRH without a mitigation, thus inducing Rowhammer bit-flips and rendering resource-limited trackers such as TRR non-secure.

\vspace{0.03in}
Emerging trackers attempt to avoid such thrashing-based attacks with deterministic tracking algorithms, which require a larger number of counters.
%requiring larger number of counters. On one hand,  
Graphene~\cite{park2020graphene} maintains activation counts using the Misra-Gries algorithm~\cite{MG}; recent proposals~\cite{kim2022mithril, ProTRR} store such counters within the DRAM logic area. 
However, the required counters  per bank increases as the TRH decreases; as shown in \cref{fig:intro}(a), Graphene and similar solutions require $\sim$5K counters per bank at TRH of 500 (680KB of SRAM per DRAM rank in DDR5~\cite{qureshi2022hydra}); such high storage overheads make these solutions impractical for in-DRAM adoption.
Alternative solutions maintain one counter per row, requiring 8K to 16K counters per bank. These counters are stored in memory~\cite{qureshi2022hydra,kim2014architectural} and require additional DRAM accesses to fetch and update the counters, leading to high worst-case performance overheads of up to 70\%~\cite{qureshi2022hydra}; storing them entirely within the DRAM array~\cite{bennett2021panopticon} requires complex redesigns of the DRAM MATs. Consequently, such deterministic trackers requiring thousands of counters per bank have been difficult to adopt in commodity DRAM.

\vspace{0.03in}

While trackerless solutions such as PARA~\cite{kim2014flipping} exist which probabilistically issue mitigations to adjacent rows on activations, incurring no storage overhead, such solutions however cannot be implemented transparently within the DRAM. They are required to be implemented only within the memory controller, as mitigative refresh commands cannot be issued by the DRAM transparently after any given activation.

\vspace{0.03in}
For an in-DRAM only solution, we seek a defense that  (a) incurs low storage  (b) has strong security (c) has low performance overheads (d) scales to low row hammer thresholds ($<1K$), and (e) is compatible with current DRAM interfaces.
%(must remain low even as the Rowhammer threshold drops to 1K or below and such a solution should still be 
Thus, we focus on low overhead mechanisms to secure
%study resource-constrained 
in-DRAM trackers (like TRR) 
%and propose thrash resistance
%with thrash-resistance
by making them thrash-resistant.
%\footnotemark{}
%for such trackers to make them secure with 
%at negligible overheads. 
%combining prior probabilistic and tracker based solutions. 
%Our goal is to provide strong security for in-DRAM trackers without additional storage and negligible slowdown.

%\TODO{}: Key idea:  prevent the tracker from thrashing using probabilistic management of trackers. Existing designs for thrash resistance in caches attempt to preserve some portion of the access stream so some get hits.  But this prevents those are not in the tracker from getting row hammer mitigations.

%\footnotetext{While thrash-resistance has been widely studied for processor caches through replacement policies like BIP~\cite{bip}, SRRIP~\cite{rrip}, etc., such solutions typically preserve a portion of the access stream in the cache to prevent thrashing. Such schemes cannot be directly used in rowhammer trackers, as activations not preserved in the tracker could then be used for Rowhammer.}

%\TODO{}: Soluution:  probabilistically sample the input request stream (PRESS). Probabilistically sampling ensures that you insert sufficient elements into the tracker without overflowing the tracker and allowing entries to be selected for mitigation. You need to sample at the sweet spot - low enough that you dont thrash the tracker, but high enough that the activationst that escaped cannot induce rowhammer.

\newpage

This paper shows that in-DRAM trackers can be made thrash-resistant by enhancing the tracker management policies, which heavily influence the tracker's security properties. Conventional tracker management policies typically consist of (a) a lookup policy (which DRAM activations should consult the tracker) (b) an update policy (how to update tracker state on hits) (c) an eviction policy (which entry to evict when it is capacity-limited), (d) an insertion policy (whether to insert entries on a miss), and finally (e) a mitigation policy (which entry to issue a mitigation for). For instance, the in-DRAM tracker in TRR~\cite{hassan2021uncovering} has a lookup policy where the activation stream is sampled at deterministic time instances, 
and mitigation/eviction policies, though not documented, that are also likely deterministic (as attack patterns aligned to mitigation instances have shown better success in deterministically evicting entries~\cite{de2021smash}). 
%and a mitigation policy that is activation-count-based (the eviction-policy is not documented but expected to be deterministic since access patterns synchronized to mitigation instances ). 
Such deterministic policies are the root cause of vulnerability to thrashing-based attacks.
%Trackers using Misra-Gries summaries like Graphene~\cite{park2020graphene} have an insertion policy of inserting all missing activations into the tracker, and frequency (activation-count) based eviction and mitigation policies. 

Based on this observation, we propose \emph{\underline{PRO}babilistic \underline{T}rack\underline{E}r m\underline{A}nagement policie\underline{S} (\PRESS{})} to enable thrash-resistant in-DRAM trackers. 
We observe that non-determinism can be best introduced by varying the insertions and evictions, rather than the mitigation policy. 
To that end, \PRESS{} proposes (i) probabilistic sampling to minimize the number of insertions into the tracker and (ii) randomize replacement to ensure that a diversity of rows are retained in the tracker.

We systematically study two different sampling approaches to minimize thrashing. 
First, thrashing can be prevented by modifying the tracker insertion policy to perform \emph{Probabilistic Miss Stream Sampling (PMSS)}, where only a subset of the misses are inserted into the tracker while the others are bypassed (similar to the Bimodal Insertion Policy (BIP)~\cite{bip} for caches). 
A similar scheme was proposed recently by Samsung's probabilistic tracker, DSAC~\cite{DSAC}\footnotemark{}. 
However, we show that sampling the miss stream is ineffective as the periodic insertions continue to thrash the small tracker. 
%This is because the miss stream reflects the total working set of the attack pattern.
Alternatively, we can modify the tracker lookup policy to perform \emph{Probabilistic Request Stream Sampling (PRSS)} where only a subset of the request stream looks up the tracker while the rest are bypassed. 
We observe that PRSS performs significantly better than PMSS (or DSAC) 
because sampling the request stream results in a working set smaller than the tracker capacity. 
%This significantly reduces the chance of evictions on insertions and 
This ensures that inserted rows get mitigated without the adversary being able to thrash them, 
and thus provides much stronger protection against thrashing-based attacks.
%This ensures that the adversary cannot avoid mitigations for its hammered rows by inducing thrashing-based evictions.
%because the sampled request stream presents a smaller effective working set at insertion time. Furthermore, when the insertion rate matches the %mitigation rate, the sampled requests receive subsequent mitigations. 
Consequently, \PRESS{} adopts  PRSS for its sampling-based thrash-resistance. 

\footnotetext{~\PRESS{} aligns closely with recent work by Samsung (i.e., DSAC~\cite{DSAC}) and SK Hynix~\cite{HynixRH} who also employ non-determinism through PRNG based Rowhammer trackers. Samsung's DSAC uses miss-sampling (like PMSS), making it less thrash resistant; in \cref{sec:max_dist} we find that it is insecure for Blacksmith-like attacks~\cite{jattke2021blacksmith}. 
Hynix provides insufficient details for us to provide a reasonable security comparison.}
%Hynix advocates request sampling, but provides no implementation or evaluation details to reason about its security.} 

However, thrash resistance alone is not sufficient. If the sampling rate is not carefully selected, too low of a sampling rate can allow requests that escape sampling to be used for rowhammer, and too high of a sampling rate can lead to thrashing. 
Hence, we derive the optimal sampling rate where the tracker just begins to thrash, when the insertion probability equals the mitigation probability of the tracker (1 per tREFI). To prevent such thrashing from being exploitable, \PRESS{} proposes a random replacement policy to ensure a diversity of rows are stored in the tracker, and thus the frequency-based mitigations are to diverse rows, as shown in \cref{fig:intro}(c). 
The tracker sampling decisions are based on a pseudo-random number generator (PRNG) with a secret seed (which is periodically changed), so the attacker cannot align with and exploit sampling instances~\cite{jattke2021blacksmith}.

We evaluate \PRESS{} on 500 uniform and non-uniform attack patterns based on TRRespass\cite{frigo2020trrespass} and Blacksmith\cite{jattke2021blacksmith}. 
Across all patterns, we measure \emph{maximum disturbance}, the maximum activations any row receives before a mitigation. With 1 mitigation per tREFI, \PRESS{} limits the maximum disturbance to 2.1K, making it a suitable defense for current (LP)DDR4 DRAM with TRH of 4.9K to 9K~\cite{kim2020revisitingRH,HalfDouble}. 

%DRAM like DDR4 and LPDDR4 that have a row hammer threshold of 4.9K to 9K~\cite{kim2020revisitingRH,HalfDouble}. 

We also co-design \PRESS{} with RFM (a feature in DDR5), which allows extra mitigation opportunities to the DRAM. We observe that a baseline deterministic tracker (like TRR), benefits minimally from additional RFM mitigations, as it is easily thrashed even in the shorter period between mitigations; it thus suffers a maximum disturbance of 73K to 64K activations with 1 to 8 mitigations per tREFI. Whereas, \PRESS{} with RFM significantly benefits from additional mitigations: 
with 4 mitigations per tREFI, \PRESS{} limits the maximum disturbance to 600, which reduces to 300 with 8 mitigations per tREFI. This makes \PRESS{} a scalable defense even as TRH drops to 1K or 500 in the future.
%with DDR5 and beyond.

Compared to recent probabilistic designs, our in-DRAM solution \PRESS{} achieves a maximum disturbance that is 15\% lower than  PARA~\cite{kim2014flipping}, a trackerless solution in the memory controller, and $19X$ lower than Samsung's DSAC~\cite{DSAC}, a probabilistic in-DRAM tracker at equivalent mitigation costs. 
%Unlike PARA, which is an in-memory controller design that requires knowledge of neighboring rows in DRAM. 

\ignore{
In comparison, DSAC~\cite{DSAC}, which uses a probabilistic insertion policy for its tracker, with a decreasing sampling probability has a maximum disturbance as high as \redx{32K}; this is mainly because non-uniform attack patterns can exploit its low sampling probabilities and continue to induce Rowhammer, as we explain in Section-\redx{X}.
}

%\TODO{}: Can we do better?  And we find that we can increase the number of mitigations issued per TREFI, for DDR5 and beyond.  This brings the maximum distrubance down further (closer to 250) at the cost of additional command bandwidth overhead to DRAM.  This makes the system secure against Rowhammer till TRH$=$500 while incurring 3\% slowdown.

%\TODO{}: Compared to a recent work DSAC that also proposed probabilsitic sampling, we lower the maximum disturbance by orders of magnitude. The key reason being that it focuses on thrash resistance (protecting against TRRespass) while allowing addresses not inserted into the tracker to continue hammering the DRAM (like Blacksmith). 

Overall, this paper makes the following contributions:
\begin{enumerate}
    %\item We systematically analyze the benefits of the different sources of non-determinism in Rowhammer trackers, including probabilistic sampling based insertion and eviction policies. 
    \item We show that effective tracker management can enable in-DRAM trackers to be secure against row hammer attacks at Rowhammer thresholds of 1K or below.
    \item We propose ~\PRESS{}, a probabilistic tracker management policy, that uses request stream sampling and random replacement to prevent tracker thrashing.%that uses, that reduces max-disturbance by 16x and 8x compared to prior approaches like TRR and DSAC, at negligible slowdown.
    \item We demonstrate how a recent probabilistic tracker, Samsung's DSAC~\cite{DSAC}, is insecure and incurs a max disturbance beyond TRH, and 19X higher than \PRESS{}.
    \item We co-design \PRESS{} with RFM on DDR5 systems to show \PRESS{} can significantly lower the max disturbance below TRH of 1K and 500, unlike deterministic trackers (like TRR) in commodity DRAM which benefit minimally from extra RFM mitigations.
    %the number of mitigations per tREFI for further protection at ultra-low row hammer thresholds.  
    
\end{enumerate}

To our knowledge, this is the first work to systematically analyze the relationship between a tracker's management policies and its susceptibility to rowhammer. 
Our Gem5 evaluations with SPEC CPU2017 workloads show that the average performance impact is less   than 1\% for a TRH of 1K and 3\%  for TRH of 500. Our in-DRAM implementation incurs negligible storage (less than 3KB per DRAM rank) that is similar to current TRR implementations.

\section{Background and Motivation}
%In this section, we first describe the DRAM Rowhammer effects, DRAM architecture, and recent mitigations and their drawbacks.

\subsection{DRAM Architecture and Parameters.}

\textcolor{black}{To access data from DRAM,} the memory controller \textcolor{black}{first} issues an activate (ACT) for a DRAM row \textcolor{black}{to} read it into a row-buffer, the 64B data is \textcolor{black}{then} transferred to the bus, \textcolor{black}{and} then the row may be closed and pre-charged (PRE). 
For reliable operation, the memory controller ensures a minimum time gap of \texttt{tRC} between two successive ACTs to a bank. \textcolor{black}{The} charge of a DRAM row must \textcolor{black}{also} be refreshed periodically every tREFW time period.
For this, the memory controller issues a REF command every tREFI period to refresh a subset of rows. There are 8192 REFs issued every tREFW period to cover all DRAM rows. \cref{table:Params} provides these timing parameters.

\begin{table}[!htb]
  \centering
  \vspace{-0.1in}
  \caption{DRAM Parameters (from Micron DDR4 datasheet~\cite{micron_ddr4})}
  \vspace{-0.1in}
  \begin{footnotesize}
  \label{table:Params}
  \begin{tabular}{lcc}
    \hline
    \textbf{Parameter} & \textbf{Explanation} & \textbf{Value} \\ \hline
    \rule{0pt}{0.8\normalbaselineskip}tREFW     & Refresh Period & 64 ms \\ %\hline
    tREFI     & Time between successive REF Commands & 7800 ns  \\ %\hline
    tRFC      & Execution Time for REF Command & 350 ns  \\ %\hline    
    tRC       & Time between successive ACTs to a bank & 45 ns \\ %\hline \hline  
    ACTs-per-tREFI & ( tREFI - tRFC )~/~tRC & 165 \\ \hline 
  \end{tabular}
  \vspace{-0.1in}
  \end{footnotesize}
\end{table}

\subsection{DRAM Rowhammer Attacks}
Kim et al.~\cite{kim2014flipping} discovered that rapid activations of a DRAM row (called Rowhammer) can cause charge leakage and bit flips in neighboring rows. 
The heavily activated row is called an ``aggressor'' row and the neighboring row with a bit-flip is called the ``victim'' row. These bit flips can occur up to distance of 2 rows from the aggressor rows (also known as the ``blast-radius'')~\cite{HalfDouble}.
The minimum number of activations to an aggressor row to cause a bit-flip in a victim row is called the ``Rowhammer threshold'' (TRH).
With DRAM scaling, TRH has dropped significantly from 139K for DDR3 in 2014~\cite{kim2014flipping} to 10K for DDR4~\cite{kim2020revisitingRH} and just 4.8K--9K for LPDDR4 in 2020~\cite{HalfDouble,kim2020revisitingRH}.
\textcolor{black}{(see \cref{table:RHT})}.

Rowhammer is not only a reliability issue but has also become a critical security threat. Numerous exploits have shown that Rowhammer bit-flips in page-tables or sensitive binaries can be used by attackers to escalate to kernel-level privileges~\cite{seaborn2015exploiting,gruss2018another,frigo2020trrespass,zhang2020pthammer}, or the data-dependent nature of the bit-flips can  be used to leak confidential data~\cite{kwong2020rambleed}. Thus, it is imperative to reduce the risk of Rowhammer attacks.

%Moreover, in recent attacks~\cite{HalfDouble}, bit-flips have been observed in victim rows at distance 2 from the aggressor rows. We denote this distance as the  ``blast-radius'' of the attack.

\ignore{
\begin{table}[!t]
  \centering
  \begin{footnotesize}
  \caption{\textcolor{black}{Rowhammer Threshold Over Time}}
  \label{table:RHT}
  \begin{tabular}{cc}
    \hline
    \textbf{DRAM Generation} & \textbf{RH-Threshold} \\ \hline %\hline
    \rule{0pt}{0.8\normalbaselineskip}DDR3 (old)     & 139K ~\cite{kim2014flipping} \\ %\hline
    DDR3 (new)     & 22.4K ~\cite{kim2020revisitingRH} \\ %\hline
    DDR4 (old)     & 17.5K ~\cite{kim2020revisitingRH} \\ %\hline    
    DDR4 (new)     & 10K ~\cite{kim2020revisitingRH} \\ %\hline \hline  
    LPDDR4 (old)   & 16.8K ~\cite{kim2020revisitingRH} \\ %\hline 
    LPDDR4 (new)   & 4.8K ~\cite{kim2020revisitingRH} - 9K~\cite{HalfDouble} \\ \hline   
  \end{tabular}
  \end{footnotesize}
\end{table}
}

\begin{table}[htb]
  \centering
  \begin{footnotesize}
    \vspace{-0.1in}
  \caption{\textcolor{black}{Rowhammer Threshold Over Time}}
 \vspace{-0.1in}
  \label{table:RHT}
  \begin{tabular}{cc}
    \hline
    \textcolor{black}{\textbf{DRAM Generation}} & \textcolor{black}{\textbf{RH-Threshold}} \\ \hline %\hline
    \rule{0pt}{0.8\normalbaselineskip}\textcolor{black}{DDR3}     & \textcolor{black}{22.4K~\cite{kim2020revisitingRH} - 139K~\cite{kim2014flipping}} \\ %\hline
    \textcolor{black}{DDR4}     & \textcolor{black}{10K~\cite{kim2020revisitingRH} - 17.5K~\cite{kim2020revisitingRH}} \\ %\hline    
    \textcolor{black}{LPDDR4}   & \textcolor{black}{4.8K ~\cite{kim2020revisitingRH} - 9K~\cite{HalfDouble}} \\ \hline   
  \end{tabular}
  \vspace{-0.2in}
  \end{footnotesize}
\end{table}

\subsection{Threat Model}
Our threat model assumes an attacker with native execution capability, that can issue memory requests for arbitrary addresses. 
We assume the attacker knows the defense algorithm, but does not have physical access to the memory controller or DRAM to learn any secret information stored inside DRAM (e.g., seed used for random-number generator).
Our defense aims to prevent all known forms of Rowhammer attacks, including TRRespass~\cite{frigo2020trrespass} and Blacksmith~\cite{jattke2021blacksmith}, that attempt to fool the tracker or Half-Double~\cite{HalfDouble} which fools mitigative refresh. 
The recent RowPress~\cite{rowpress} attack is out-of-scope, since its effects are orthogonal to Rowhammer and can be mitigated with a paging policy that limits the time a row is kept open.%time for which a row is kept open at minimal overhead.

\subsection{Targeted Row Refresh (TRR) Mitigation in DDR4}
DDR4 modules support in-DRAM mitigation against Rowhammer called Targeted Row Refresh (TRR)~\cite{frigo2020trrespass}. 
TRR maintains an aggressor row tracker within each DRAM bank. Each tracker entry holds the DRAM row number and a frequency counter.
When the memory controller issues a REF command every tREFI period, the DRAM issues a mitigation for the highest activated row in the tracker.
The mitigation involves refreshing the neighboring victim rows (in a given blast radius) during tRFC in the background of a REF command.
%without any performance overheads.

\textbf{Recent Attacks.} Recent studies~\cite{hassan2021uncovering,frigo2020trrespass,jattke2021blacksmith} have shown that trackers typically contain less than 32 entries per bank due to DRAM storage constraints. The \textit{TRRespass}~\cite{frigo2020trrespass} attack exploited this limited storage by uniformly hammering a large number of rows beyond the tracker capacity, to thrash the tracker, and evict resident entries. As a result, the hammered rows that are evicted escape a mitigation, and bit-flips can be induced by fooling the tracker.
Recent work~\cite{jattke2021blacksmith} also observed that some tracker implementations may perform deterministic temporal sampling   of activations for insertion into the tracker. The \textit{Blacksmith} attack~\cite{jattke2021blacksmith} exploited this to flip bits by aligning its hammering pattern to escape sampling and increasing the intensity of hammering for these rows. 

\begin{comment}
\aj{can we snip this para?}In addition to these attacks that exploited the tracker, a recent \textit{Half-Double} attack exploited mitigations. This attack exploited the fact that frequent mitigative refreshes to the neighbors can themselves be used for hammering, and in turn induce bit-flips in distance-2 neighbors. To mitigate this, the mitigative refreshes can be extended to cover 2 rows above and 2 rows below the aggressor (cover a blast-radius of 2).
\end{comment}

\textbf{Refresh Management (RFM).} DDR5  includes a new feature, RFM, \textcolor{black}{where} memory controller \textcolor{black}{can} issue additional mitigations (RFM commands) to the DRAM when the activations per bank crosses a threshold. 
However, the actual row to be mitigated still depends on the in-DRAM tracker implementation. Consequently, if a vulnerable tracker like TRR~\cite{frigo2020trrespass,jattke2021blacksmith} can be fooled (by thrashing or escaping sampling) between mitigations, it can allow existing attacks to continue despite additional RFM mitigations, as we shown in ~\cref{sec:max_dist}.

\begin{tcolorbox}[boxrule=1pt,left=5pt,right=5pt,top=3pt,bottom=3pt]
Resource-constrained trackers (like TRR with $<$32 entries) are susceptible to attacks exploiting thrashing-based evictions or deterministic sampling to escape insertions.
\end{tcolorbox}

\subsection{Storage Overheads of State-of-the-Art Trackers}
State-of-the-art trackers typically store a large number of entries to deterministically prevent thrashing-based attacks. 
Such trackers may be stored in SRAM, DRAM or in both SRAM and DRAM.
At one end of the spectrum, Graphene~\cite{park2020graphene} and Mithril~\cite{kim2022mithril} perform approximate counting and require fewer counters, while at the other end of the spectrum, solutions utilize one counter per row~\cite{qureshi2022hydra,kim2014architectural,bennett2021panopticon} to perform exact tracking.
Other trackers~\cite{CBT,lee2019twice,yauglikcci2021blockhammer} lie between these.

\textbf{Graphene:} Graphene~\cite{park2020graphene} uses activation counters based on Misra-Gries summaries, a solution to the frequent element problem, to identify aggressor rows from a stream of row activations in the memory controller. When any counter in the tracker crosses a predefined threshold (TRH/2), the associated row is mitigated. The counts in this tracker are always guaranteed to be greater than or equal to the actual row activation counts, so long as the number of tracker entries is greater than  ACTs-per-tREFW / (TRH/2). 
For TRH of 500, this requires 5440 counters per DRAM bank (87K counters per rank) consuming 240 KB SRAM per DRAM rank; additionally,
%to facilitate lookup and update, 
this must be organized as a 5400-entry CAM per bank, which may be beyond practical capabilities~\cite{qureshi2022hydra}. 

\textbf{Mithril:} Mithril~\cite{kim2022mithril} stores a similar Misra-Gries tracker within DRAM, and co-designs it with RFM to allow a smaller tracker size when the mitigations are issued at a higher frequency (with lower $RFM_{TH}$).
Unfortunately, Mithril also faces a similar problem that at low thresholds of 500, to keep performance overheads low, it requires a tracker of few thousand counters per DRAM bank, which is impractical given that deployed in-DRAM defenses only have tens of counters.  

\textbf{Hydra:} Hydra~\cite{qureshi2022hydra} and CRA~\cite{kim2014architectural} store one counter per DRAM row, in a reserved portion of the DRAM. 
A 4GB DRAM Rank (with 8KB rows) requires 512K counters per rank. 
Unfortunately, Hydra faces performance overheads due to additional DRAM accesses to fetch and update the counters. 
Although these solutions deploy complex caching and filtering mechanisms on-chip within the memory controller  to avoid extra DRAM accesses, recent work~\cite{qureshi2022hydra} shows that the average slowdown for CRA is 25\%, and for pathological workloads where Hydra's filtering may be ineffective, the worst-case slowdown can be as high as 70\%.  

\textbf{Panopticon:} Panopticon~\cite{bennett2021panopticon} and PRHT from Hynix~\cite{HynixRH} propose per-row counters stored within the DRAM array. This requires 512K counters per DRAM rank, as before, which may be updated without additional DRAM accesses.
However, such solutions require a significant redesign of the DRAM MATs, and additional logic to store and update these counters in the background of a DRAM ACT, which can additionally impact DRAM timings.
The required complex redesign of the DRAM arrays makes such solutions less desirable.
%\TODO{Table with tracker-based solutions. Make this list more exhaustive.}

\begin{tcolorbox}[boxrule=1pt,left=5pt,right=5pt,top=3pt,bottom=3pt]
State-of-the-art trackers require large storage overhead (87K -- 512K counters per DRAM rank) or require significant SRAM~/~DRAM changes making them undesirable.
%for thrash protection and incur significant SRAM storage, or complex SRAM and DRAM changes, or suffer high slowdowns, which makes them undesirable.
\end{tcolorbox}

\subsection{Challenge with Existing Probabilistic Mitigations}
PARA~\cite{kim2014flipping} or PRA~\cite{kim2014architectural} are trackerless probabilistic solutions that issue a neighboring row activation from the memory controller with a probability \textit{p} on each DRAM activation.
Such solutions are trackerless, so they incurs no storage overheads. 
Unfortunately, this cannot be implemented transparently within the DRAM as  only the memory controller can issue additional activation commands at arbitrary time instances to specific rows; thus, such solutions must be implemented within the memory controller.
This also means that the memory controller requires the knowledge of adjacent rows, which is not currently exposed by DRAM to the memory controller.
Prior works~\cite{PROHIT,MRLOC} have attempted to minimize mitigations in PARA, but at the cost of lowering the security.

\textcolor{black}{Recent proposals like PRoHIT~\cite{PROHIT}} and Samsung's DSAC~\cite{DSAC} employ stochastic replacement in resource constrained in-DRAM trackers, to filter out decoy rows in TRRespass attacks. DSAC achieves this by modifying the tracker insertion policy to progressively reduce the probability of eviction from the tracker on misses as the minimum count in the tracker increases. \textcolor{black}{PRoHIT uses two tracker tables (hot table and cold table) and probabilistically promotes entries from the cold to hot table to limit thrashing.}  
While such designs provide security against TRRespass attack, \textcolor{black}{they are vulnerable to newer attacks: e.g.,} DSAC's consistent reduction in insertion probability allows rows that escape sampling %towards the end of a tREFI period 
\textcolor{black}{to still be hammered} (like in the Blacksmith attack~\cite{jattke2021blacksmith}). 
%Our evaluations in \redx{Section-X} show that DSAC configured to mitigate a TRH of 8K can allow rows to be activated for over 16,000 activations before a mitigation is issued under
In \cref{sec:max_dist}, we show that both PRoHIT and DSAC are insecure against Blacksmith-like attack patterns, which were not evaluated in the original submissions. 
%Thus DSAC itself is not secure.
%\footnote{Hynix~\cite{HynixRH} similarly showcased the fabrication of a DRAM chip with a Rowhammer tracker and an in-DRAM pseudo-random number generator, but with no implementation details or evaluations its security is unknown.}

\subsection{Goal: Low Cost, Scalable, and Secure Trackers}
Ideally, we desire a secure Rowhammer mitigation solution that incurs low performance and storage overheads and is compatible with existing DRAM protocols to be easily adoptable by current and future DRAM. Furthermore, we desire a scalable solution as Rowhammer thresholds drop to 500 (by 10x compared to 2020 levels). To that end, we study storage-limited trackers (like TRR) and redesign them to be secure by systematically enabling them to be thrash-resistant.

\begin{comment}
## -- Rowhammer thresholds are dropping. Security and reliability problem.
## -- Existing solutions - Memory Controller based (PARA, HYDRA, AQUA, Mythril,etc.) or In-DRAM tracker based (Graphene, etc.)

\end{comment}
\section{Methodology}\label{sec:analytical_methods}
%\noindent \TODO{:describe methodology of analytical evaluations?}
% Analytical simulator.
% Atleast 64ms 
% Aligned and Non-Aligned Patterns.
% TRRespass like Uniform Patterns with Different Pattern Lengths.
% Blacksmith like Non-Uniform patterns varying intensity, frequency, phase.
To analyze the thrash-resistance properties of trackers, we use an empirical methodology based on access patterns derived from recent row hammer attacks, like TRRespass~\cite{frigo2020trrespass}, Blacksmith~\cite{jattke2021blacksmith}, and SMASH~\cite{de2021smash}. We develop a trace-driven Rowhammer simulator which models our trackers and uses DRAM activation traces of attacks as input. 

\textbf{Attack Patterns:} We use \emph{uniform}~\cite{frigo2020trrespass} and \emph{non-uniform}~\cite{jattke2021blacksmith} attack  patterns (see \cref{table:AttackPatterns}).  A uniform attack pattern resembles a TRRespass attack~\cite{frigo2020trrespass}, consisting of a cyclical reference to a set of target rows $r$ with length $j$ (larger than the tracker capacity). A non-uniform pattern is like Blacksmith~\cite{jattke2021blacksmith}, where activations occur to a set of target rows $r$ with length $j$ with a higher intensity ($X$), compared to a second set of rows $d$ with length $k$, with an intensity of 1. 
The difference in intensity ($X$) and sequence lengths ($j$, $k$) achieves the effect of varying intensity, phase, and frequency like in Blacksmith~\cite{jattke2021blacksmith}. 
%The target rows are repeatedly activated with intensity $X$ in a cyclical pattern followed by activations to the decoy rows. The primary goal of the decoy accesses are to discard the frequently activated target rows from the tracker. 
We evaluate 10 uniform activation patterns for different values of $j$ (see Table~\ref{table:AttackPatterns}).  For non-uniform activation patterns, we evaluate ten different values for $j$, four different values for $X$, and six different values for $k$ for a total of 240 non-uniform patterns. Recent attacks~\cite{de2021smash} show that aligning activation patterns to tREFI can further increase the success of row hammer attacks. Thus, we evaluate aligned and unaligned versions of our 250 attack patterns for a total of 500 patterns. 
These patterns cover footprints of $2-220$ unique addresses.

\begin{table}[!t]
  \centering
  \vspace{-0.1in}
  \caption{Attack Pattern of Activations Under Study}
  \begin{footnotesize}
  \label{table:AttackPatterns}
  \begin{tabular}{lcc}
    \hline
    \textbf{Type} & \textbf{Pattern} & \textbf{Parameter Sweep} \\ \hline
    \rule{0pt}{1\normalbaselineskip}Uniform     & $(r_1,...,r_j)^N$ & $j=2,4,8,16,20,$\\ %\hline
    & & $32,40,80,120,140$ \\[0.1in]
%    & & \\
    Non-Uniform     & $[(r_1,...,r_j)^X,(d_1,..., d_k)]^N$ & $j=2,4,8,16,20,$  \\ %\hline
    & & $32,40,80,120,140$ \\
    & & $X=2,3,4,5$\\
    & & $k=5,10,20,32,40,80$\\
  \end{tabular}
  \vspace{-0.2in}
  \end{footnotesize}
\end{table}

\textbf{Simulator:} We use a trace-based simulator modeling an aggressor-row tracker assuming DRAM parameters listed in \cref{table:Params}. We assume a baseline 16-entry fully-associative tracker. We run our 500 attack patterns through the simulated tracker. 
%(the attack patterns are up to $14x$ larger than our baseline tracker). 
For our baseline tracker, 
on hits, the frequency counter of the associated entry is incremented. On every miss, a new entry is inserted into the tracker with the frequency counter set to zero. In the event the tracker is full, the Least Frequently Used (LFU) entry in the tracker is evicted. 
At tREFI intervals, the Most Frequently Used (MFU) entry is selected for mitigation after which the entry is invalidated. The attack patterns are continually repeated for a value $N$ that covers one refresh interval. Like prior work~\cite{DSAC}, we report ``maximum disturbance'' which is the maximum activations received by any row in the attack pattern before being refreshed.
% and report the average \emph{maximum disturbance} metric across these 10 runs.   
\section{Probabilistic Tracker Management}
We first formalize the management policies of a typical aggressor-row tracker. We then systematically study probabilistic tracker management policies for thrash protection.

\begin{figure}[!t]
    \centering
    \includegraphics[width=3.5in]{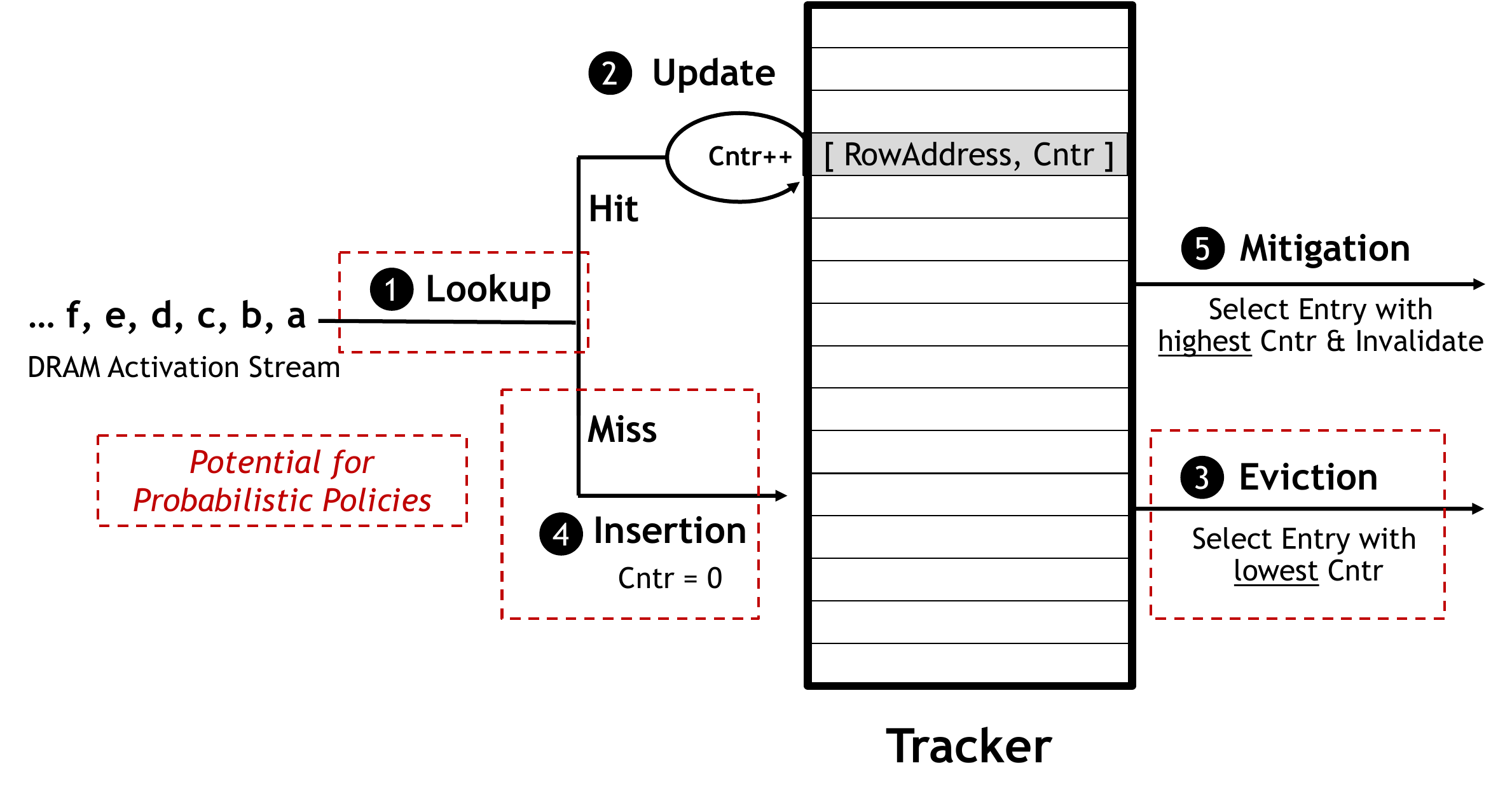}
    \vspace{-0.2 in}
    \caption{Overview of Tracker Management Policies and Potential for Probabilistic Policies for Thrash Resistance}
    \label{fig:TrackerDesign}
    \vspace{-0.1 in}
\end{figure}

\subsection{Formalizing Tracker Management Policies}
\textbf{Structure.} \cref{fig:TrackerDesign} shows a representative aggressor-row tracker typical in DRAM defenses like TRR.
%storing activation counts for recently activated DRAM rows. 
We study small trackers (e.g. 16 entry) which can be fully-associative in-DRAM tables.  Each entry stores the DRAM row address and a frequency counter, which counts the hits received by the entry while resident in the tracker. 
%DRAM vendors periodically consult the Tracker to identify the most frequently referenced Tracker-entry that should be issued a row hammer mitigation.  It is highly desirable that the Tracker store frequently referenced rows from the DRAM activation stream that are susceptible to row hammer attacks. Thus, Tracker management is key to mitigating row hammer vulnerability.
Since the entries with the highest counts are selected for mitigation, a fool-proof tracker management policy that retains frequently activated rows is critical for effective mitigation.

%\subsection{State-of-the-Art Tracker Management Policies}

\textbf{Policies.} As shown in \cref{fig:TrackerDesign}, tracker management is composed of many different policies: \circled{1} Lookup Policy, \circled{2} Update Policy, \circled{3} Eviction Policy, \circled{4} Insertion Policy, and \circled{5} Mitigation Policy.
In state-of-the-art trackers (like TRR~\cite{hassan2021uncovering} or Graphene~\cite{park2020graphene}), the Lookup Policy typically consults the tracker on \emph{all (or a deterministic subset of)} DRAM activations~\cite{jattke2021blacksmith}.  On a hit, the Update Policy increments the associated counter. 
%This allows the Tracker to identify frequently referenced entries within the Tracker. 
On a miss, the Insertion Policy typically inserts \emph{all} missing entries into the tracker~\cite{park2020graphene} with a frequency count of zero. 
If the tracker is full\footnote{The insertion policy prioritizes inserting into invalid entries first.}, the Eviction Policy typically employs a Least Frequently-Used (LFU) policy where the entry with the lowest count is evicted.
%Finally, when issuing a mitigation (e.g., at tREFI intervals), the Tracker Mitigation Policy employs the Most Frequently-Used (MFU) policy where the entry with the highest access count is selected for  mitigation~\cite{park2020graphene,hassan2021UTRR}. Upon issuing the mitigation, the Tracker-entry is invalidated to allow other Tracker entries to receive mitigations. 
Finally, when issuing a mitigation, the Mitigation Policy selects the entry with highest count for mitigation~\cite{park2020graphene,hassan2021UTRR}, after which the entry is invalidated to enable other frequently accessed rows to receive mitigations. Without loss of generality, 
%for an in-DRAM tracker, 
we assume mitigations are issued at tREFI intervals and refresh neighboring rows within a blast-radius of 2 (to protect against Half-Double~\cite{HalfDouble}). 

\subsection{Probabilistic Tracker Management Policies (\PRESS{})}
%\subsection{Towards Probabilistic Tracker Management with PROTEAS}

Unfortunately, \emph{deterministic} policies in state-of-the-art trackers are susceptible to \emph{thrashing} attack patterns. Such attack patterns are designed with a \emph{working set} larger than the tracker capacity to 
%evict inserted entries
cause thrashing, or are cleverly designed to avoid insertions into the tracker exploiting the deterministic policies. 
%They take advantage of the aforementioned deterministic management policies to displace target rows from the Tracker. Consequently, the Tracker Mitigation Policy is unable to mitigate target rows since they are not Tracker resident.  
To address this, we enable thrash resistance with \emph{\underline{PRO}babilistic \underline{T}rack\underline{E}r m\underline{A}nagement policie\underline{S} (\PRESS{})}\footnote{Named after the Greek God Proteus who could change his form at will. PROTEAS probabilistically changes insertion patterns seen by the tracker.}.

%\subsection{Probabilistic Tracker Management Policies (PROTEAS)}
%PROTEAS consists of three modifications to the baseline Tracker management policy. First, we modify the baseline Tracker Replacement Policy to use \emph{random replacement}. Random replacement introduces non-determinism and avoids thrashing by allowing different rows to stay resident within the Tracker on misses. Second, we modify the Tracker Lookup Policy to sample a subset of the request stream. In doing so, we can avoid thrashing and retain different portions of the DRAM activation stream within the Tracker over the refresh interval. We refer to this as \emph{Probabilistic Request Stream Sampling (PRSS)}. Finally, we modify the Tracker Insertion Policy to insert a subset of the Tracker miss stream. Doing so also avoids thrashing and can retain different portions of the DRAM activation stream within the Tracker over the refresh interval. We refer to this as \emph{Probabilistic Miss Stream Sampling (PMSS)}. These three proposals can operate independently or in conjunction with each other. We now elaborate on each of these proposals.

PROTEAS consists of two key components to introduce non-determinism and diversity in rows resident within the tracker: probabilistic sampling of insertions to avoid thrashing and a random-replacement based Eviction Policy. Unlike conventional tracker policies, both of these mechanisms enable diversity in mitigations. We systematically explain the design of these two components and their benefits.
We retain the baseline Update and Mitigation Policy in \cref{fig:TrackerDesign} (i.e, frequency counter update on  hits and MFU-based mitigations) since we still desire that the most heavily activated rows are tracked and mitigated accurately.   

\begin{figure}[!t]
    \centering
    \includegraphics[width=3in]{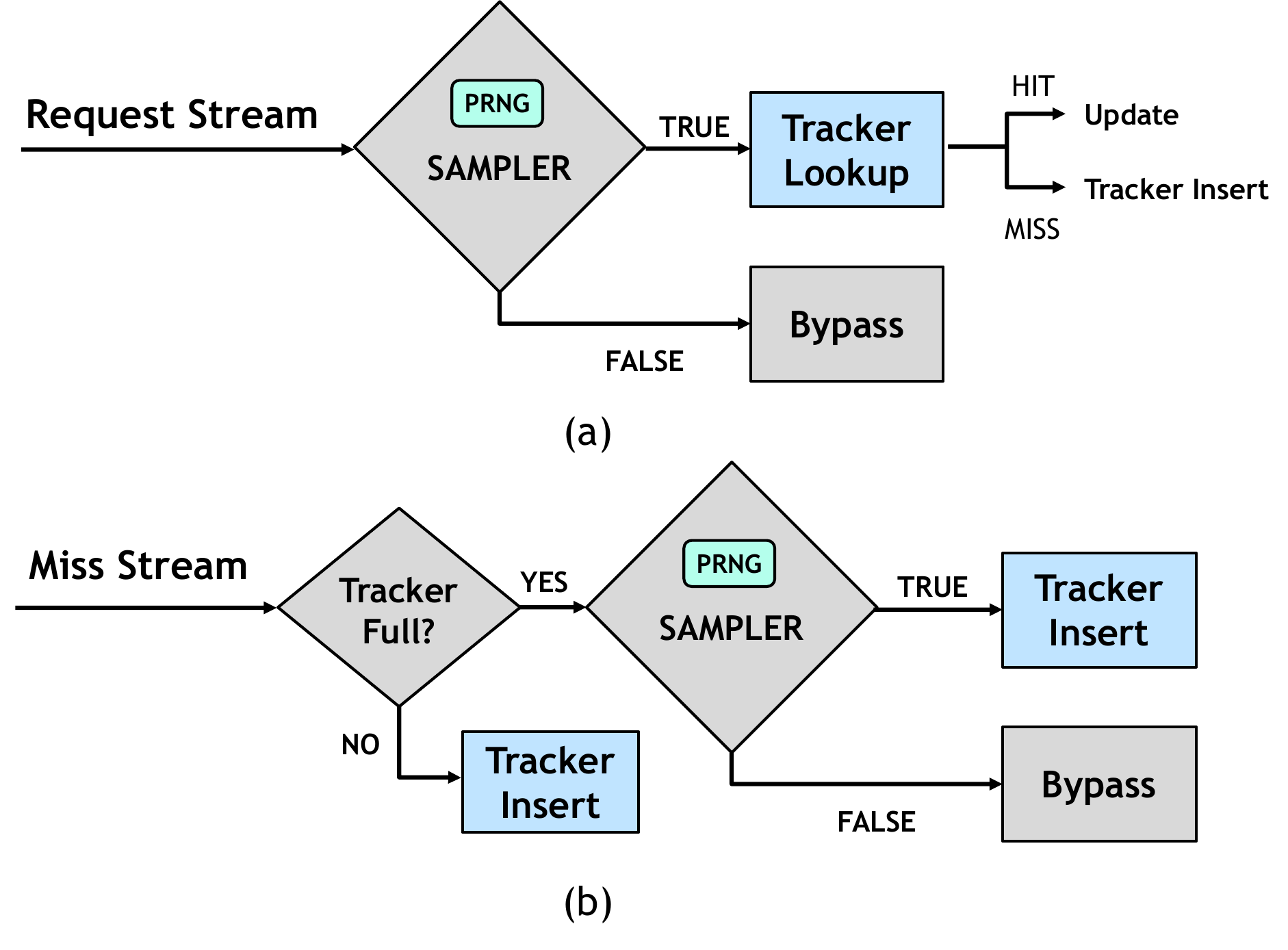}
    %\vspace{-0.15 in}
    \caption{Probabilistic Sampling of (a) Request Stream and (b) Miss Stream}
    \label{fig:PRSS_PMSS}
    \vspace{-0.1 in}
\end{figure}

\subsection{Probabilistic Sampling to Limit Tracker Thrashing}
Our goal is to prevent thrashing based attacks from being able to fool the tracker by deterministically evicting tracked entries. 
Probabilistic sampling of insertions aims to reduce the number of insertions to limit thrashing capability of the attack. At the same time, the probabilistic sampling decisions are based on a pseudo-random-number generator (PRNG), whose seed is a secret stored within the DRAM and not known to the attacker, so the attacker may not surgically avoid sampling into the tracker.
Moreover, the seed can be changed periodically, to prevent an attacker from reverse-engineering the sequence.
Intuitively, sampling to limit insertions can be implemented either at the Miss Stream or Request Stream.
We analyze both.

\textbf{Probabilistic Miss Stream Sampling (PMSS):} Typical Insertion Policies insert \emph{all} missing entries into the tracker. When the tracker is full, the Eviction Policy is used to select a suitable victim. Consequently, when the access pattern has a footprint that is larger than the tracker size, the tracker starts thrashing. Past studies have shown that thrashing can be avoided by inserting only a subset of the  miss stream and bypassing the rest~\cite{bip}. %Probabilistically choosing from the tracker miss stream allows different portions of the access stream to be tracker resident over the DRAM refresh interval.  
%This preserves a portion of the working set in the tracker, and intermittently installs new entries. 
\cref{fig:PRSS_PMSS}(b) illustrates the implementation of PMSS. On a Miss, if there is available capacity in the tracker, the missing entry is installed in the empty locations. If the tracker is full, PMSS uses a PRNG with a sampling probability \emph{p} to insert a subset of misses into the tracker. 
This preserves a portion of the working set in the tracker, and intermittently installs new entries, thus allowing new untracked rows to be tracked.
In fact, such a technique is deployed in thrash-resistant cache replacement policies like BIP~\cite{bip} and in DSAC's stochastic replacement for aggressor-row tracking.
%The sampling probability \emph{p} for PMSS must be selected appropriately to prevent thrashing while also allowing different portions of the access stream to stay resident in the tracker. For example, if \emph{p} is high, the tracker can get thrashed. With PMSS, there are no issues with tracker underutilization since the tracker Insertion policy prioritizes filling into invalid tracker entries. Again, there are no performance or correctness issues by sampling the miss stream because the tracker holds no program state.

\textbf{Probabilistic Request Stream Sampling (PRSS):} 
An alternative approach to limit thrashing is to probabilistically sample the request stream and only use a subset of the activations to consult  the tracker.  As shown in \cref{fig:PRSS_PMSS}(a), PRSS uses a sampling probability \emph{p}, based on a a PRNG, to select a subset of the requests to lookup the tracker. The key idea is that frequently accessed rows have a higher chance of being sampled.  Only the sampled requests update on hits and or can insert into the tracker on misses,  while the non-sampled requests bypass the tracker.

For both PMSS and PRSS, the sampling probability \emph{p} must be selected appropriately. If \emph{p} is high, the tracker can get thrashed; if \emph{p} is very low, the non-sampled activations can induce sufficient hammering while escaping mitigations.

\begin{figure}[!t]
    \centering
    \includegraphics[width=3in]{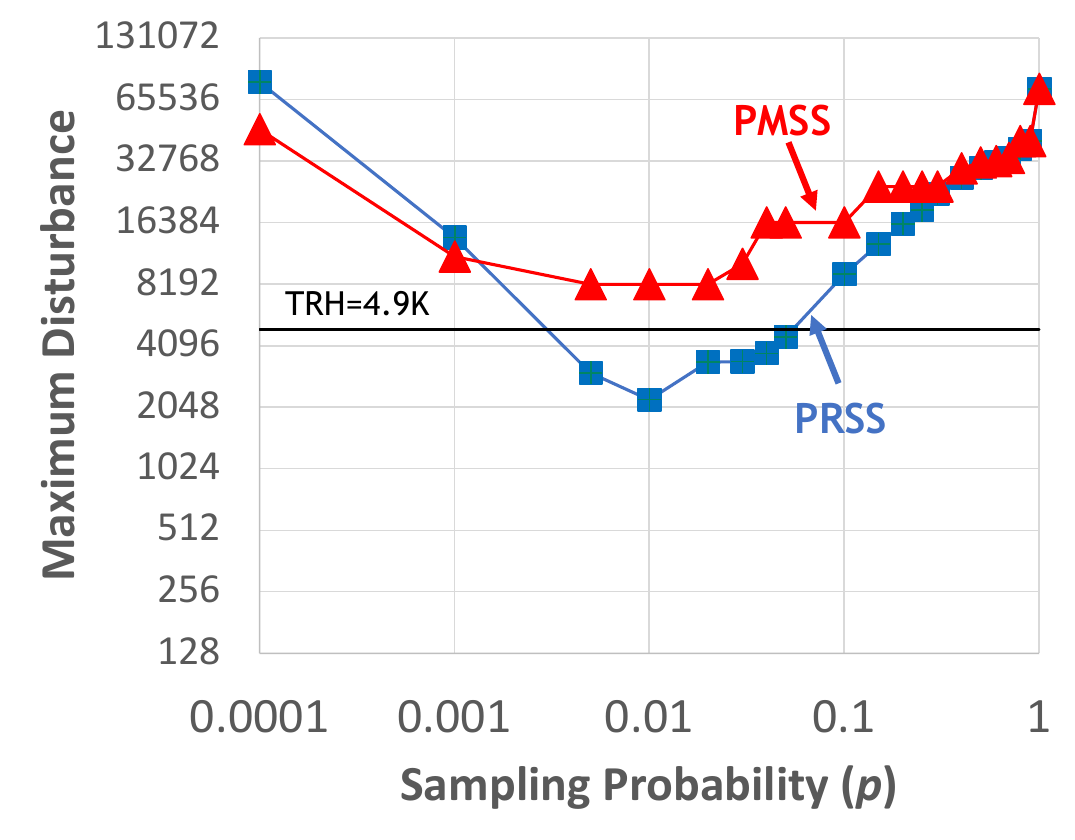}
    \vspace{-0.15 in}
    \caption{Sensitivity of PMSS and PRSS to Sampling Probability ($p$). PRSS achieves a significantly lower minima for the maximum disturbance (2.2K) compared to PMSS (5.1K), due to its more effective thrash reduction.}
    \label{fig:PRSS_PERF_Bathtub}
    \vspace{-0.1 in}
\end{figure}

\textbf{Results.} \cref{fig:PRSS_PERF_Bathtub} shows the \emph{maximum disturbance} (i.e., the maximum number of activations any DRAM row receives before a mitigation) as the sampling probability \emph{p} varies for PRSS and PMSS (using the baseline LFU replacement), across the 500 attack patterns described in \cref{sec:analytical_methods}.
%The x-axis illustrates \emph{p} while the y-axis illustrates the \emph{maximum disturbance} (i.e., the maximum number of activations any DRAM row receives before a mitigation) across 500 attack patterns described in \cref{sec:analytical_methods}.
%The x-axis uses a logarithmic scale with base ten and the y-axis uses a logarithmic scale with base two.
The baseline policy of consulting the tracker on all requests/misses ($p=1$ or 100\%) incurs a maximum disturbance exceeding 64K, indicating that the tracker can be easily thrashed. As $p$ reduces, PRSS and PMSS both reduce tracker thrashing because the maximum disturbance decreases until $p=0.01$ (1\%). However, as $p$ reduces below  1\%, maximum disturbance starts increasing again. This is because with such low sampling rates, the tracker is severely underutilized, enabling non-sampled rows to induce hammering. This suggests that the sampling rate must ensure at least one insertion per tREFI period. With a maximum of 165 activations per tREFI (see~\cref{table:Params}), ensuring at least one entry in the tracker is populated requires a sampling probability of at least 1/165 (i.e., 0.6\%).

The lowest maximum disturbance value for PRSS is 2.2K while that of PMSS is 8.1K (both at $p=0.01$ (or 1\%)). 
%This suggests that PRSS is better than PMSS when safeguarding against row hammer attacks.
PRSS significantly outperforms PMSS because of its underlying implementation.  PMSS by design ensures that the tracker stays fully occupied because invalid entries are prioritized for insertion (see~\cref{fig:PRSS_PMSS}(b)). Thus, in the steady state, whenever PMSS performs an insertion, a tracker entry must be evicted. 
%As such, PMSS favors issuing mitigations for portions of the access pattern receiving tracker hits. 
As such, PMSS still thrashes the tracker and the portion of access pattern that are not resident escape mitigation similar to the baseline. On the other hand, with PRSS, the tracker does not thrash and the occupancy is often below the maximum capacity (on average 80\% in our studies) and it performs best when the sampling rate equals the mitigation rate. Given the low chance of eviction, the sampled activations have a higher chance of receiving mitigations, and thus PRSS achieves a much lower maximum disturbance than PMSS. Thus for \PRESS{}, we choose PRSS with $p=0.01$ (1\%) as our default design.
As our tracker just begins to have infrequent evictions at this probability, we now study replacement policy.

\begin{figure}[!t]
    \centering
    \includegraphics[width=3in]{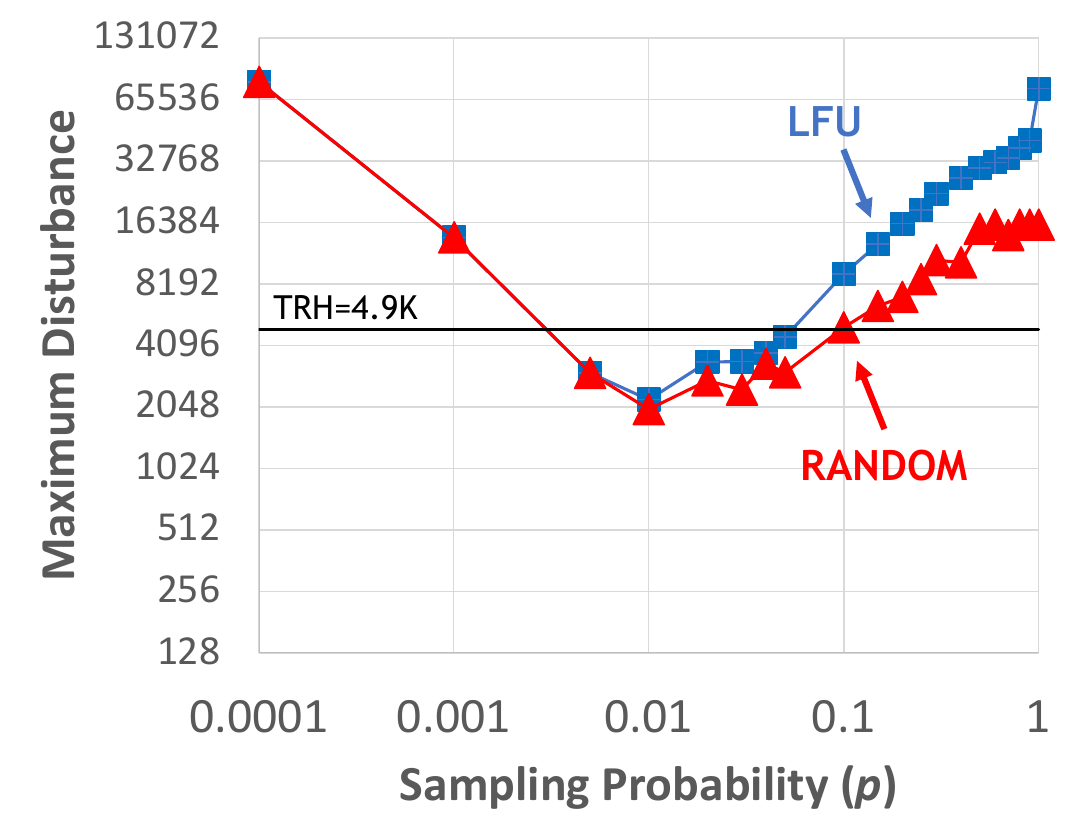}
    \vspace{-0.15 in}
    \caption{Sensitivity of PRSS to Replacement Policy. PRSS with Random Replacement achieves a 10\% lower minima for maximum disturbance (2K) compared to LFU Replacement (2.2K). }
    \label{fig:PRSS_LFU_RAND}
    \vspace{-0.2 in}
\end{figure}

\subsection{Random Replacement Policy for Diversity in Tracking}
On tracker misses, state-of-the-art trackers~\cite{park2020graphene,DSAC} employ least frequency used (LFU) replacement to select an eviction candidate. 
The goal is to retain heavily accessed rows resident in the tracker so they have the opportunity to be selected for mitigation. 
However, the disadvantage of such deterministic replacement is that it can be exploited by attack patterns to dislodge target entries deterministically.  In doing so, the target entries may be used for continued hammering while escaping mitigations (since they are not  resident in the tracker). 
%While this is a fine approach, attackers can take advantage of the deterministic frequency based replacement policy to construct access patterns to dislodge frequently accessed target rows from the Tracker. 
To address this vulnerability, we propose \emph{random replacement} on evictions. 
This ensures that evictions are unpredictable to the adversary, thus thwarting attacks that may attempt to dislodge a specific entry. Furthermore, random replacement ensures a diversity of rows are retained in the tracker allowing diversity in mitigations. Note that we only modify the Eviction Policy to use random replacement. We retain and continue to update the frequency counters used by the MFU-based Mitigation Policy, which selects the entry with the highest counter for mitigation.

\textbf{Results.} \cref{fig:PRSS_LFU_RAND} shows the maximum disturbance for PRSS with LFU and Random replacement. 
We observe that PRSS with Random replacement is almost always better than LFU replacement. In fact, it achieves a 10\% lower minima for the maximum disturbance (2K) compared to LFU based replacement (2.2K). This is because random replacement further avoids thrashing from being easily exploitable by introducing non-determinism into the tracker replacement behavior.
%prevents deterministic evictions of specific entries under thrashing attacks.

\subsection{Putting it all together: \PRESS{}}
\begin{figure}[t]
    \centering
    \includegraphics[width=3.5in]{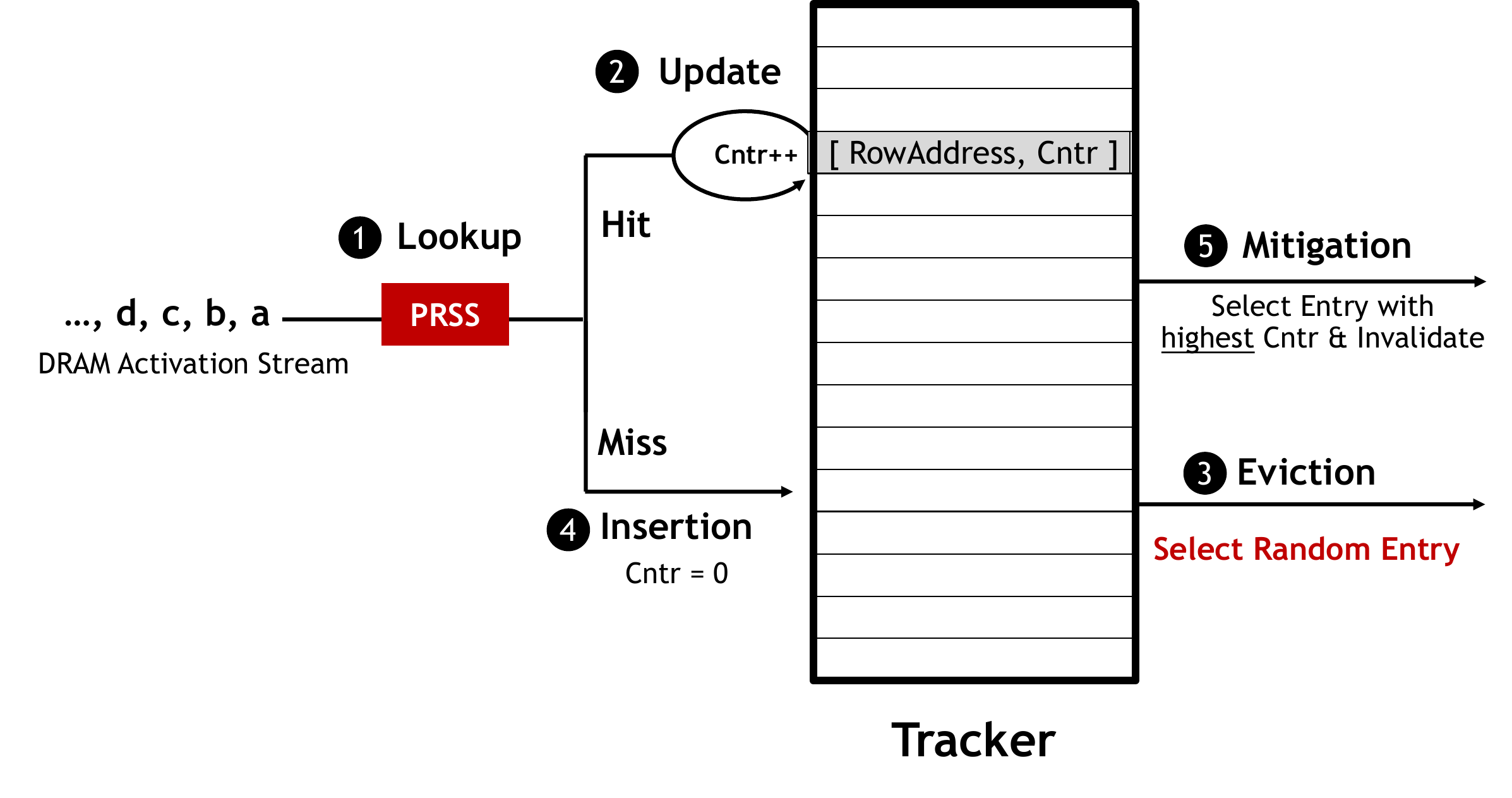}
    \vspace{-0.1 in}
    \caption{Overview of Probabilistic Tracker Management Policies (PROTEAS)}
    \label{fig:ProbTrackerDesign}
    \vspace{-0.1 in}
\end{figure}

Based on \cref{fig:PRSS_LFU_RAND}, 
\PRESS{} uses PRSS with random replacement as the default design choice. \cref{fig:ProbTrackerDesign} shows the complete design of \PRESS{}, with the PRSS and random replacement enhancements. As illustrated in the figure, ~\PRESS{} requires minimal changes to the baseline tracker design. We introduce only two pseudo-random number generators (PRNG): one for PRSS and one for random replacement. \cref{sec:design_overhead} discusses the design overhead of these modifications. 

%Random replacement ensures that a diversity of rows are retained within the tracker thwarting attack access patterns that attempt to lodge or dislodge rows from within the tracker. Our proposal only modifies the tracker Eviction Policy to random replacement and still retains the per tracker-entry frequency counters used by the tracker Mitigation Policy.

\subsection{Sizing the Probabilistic Tracker}
\begin{figure}[!t]
    \centering
    \includegraphics[width=3in]{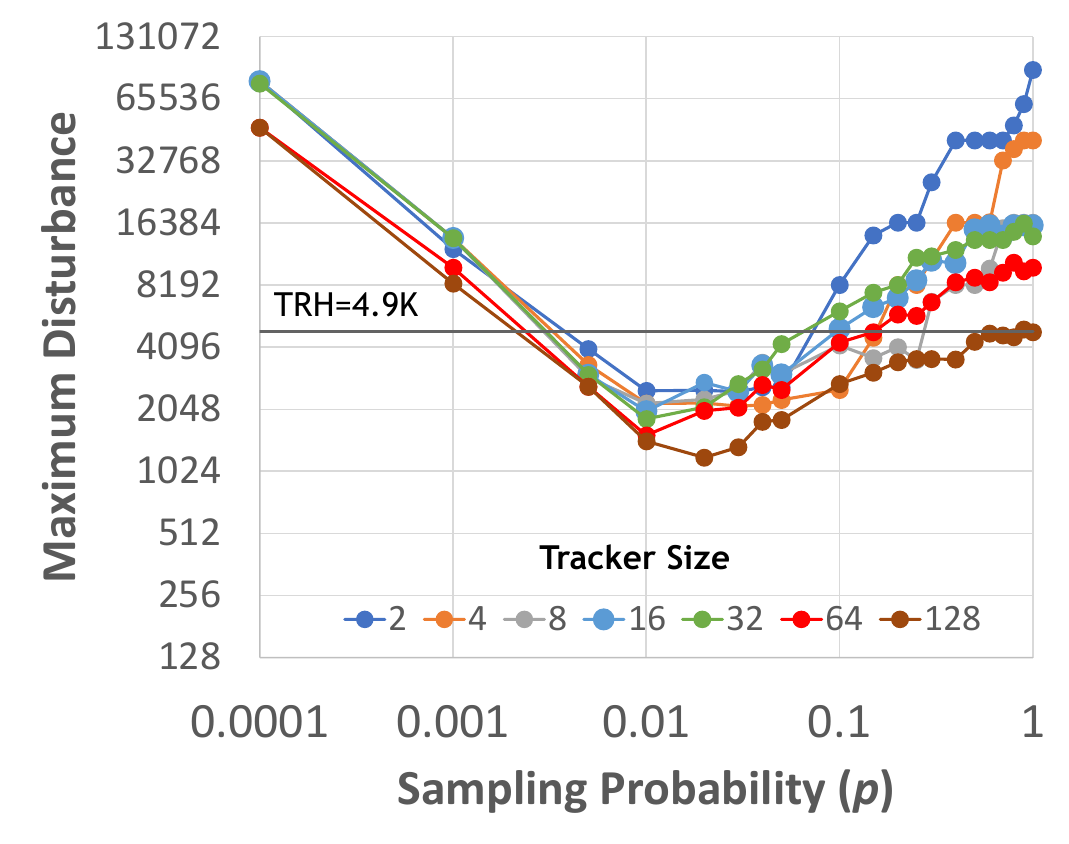}
    \vspace{-0.1 in}
    \caption{Sensitivity of PROTEAS to Tracker Size (varies from 2 to 64 counters)}
    \label{fig:TrackerSize}
    \vspace{-0.1 in}
\end{figure}

\cref{fig:TrackerSize} illustrates the sensitivity of \PRESS{} to tracker size as it varies from 2 to 128 counters. Across tracker sizes, the lowest value for the maximum disturbance is achieved close to $p=0.01$ (1\%), at which point the tracker insertion rate (approximately 1.6 per tREFI) is close to the mitigation rate (1 per tREFI). 
As tracker size increases from 2 to 128, the maximum disturbance decreases. 
For a 16-entry tracker, the maximum disturbance (at $p=0.01$) is 2K.
As the tracker size increases to 32, 64, and 128, the maximum disturbance decreases to 1834, 1538, and 1431 respectively. This is because the extra tracker capacity makes the attacker take longer to thrash the tracker. Thus, the chance that a resident entry receives a mitigation before it is evicted increases thereby reducing the maximum disturbance for an attack pattern.
On the other hand, as tracker size decreases from 16 to 4 and 2, the maximum disturbance increases as high as 2.5K. Out of an abundance of caution, we choose the default tracker size of 16 for \PRESS{} (at $p=0.01$), where the maximum disturbance of 2K is less than half the current TRH of 4.9K.

\section{Combining \PRESS{} with RFM for Scalability}
\label{sec:RFM}

Thus far, we studied \PRESS{} with one mitigation per tREFI, which is issued at the time of the regular refresh during tRFC.
Such a design is appropriate for LPDDR4 and DDR4 (and HBM2) memories, where the opportunity for issuing mitigations is only on DRAM refresh commands. 
Given the limited tRFC, it is not practical to issue mitigations for more than one aggressor row per tREFI which involves refreshing victim rows within a blast radius of 2 or 4.

In DDR5 and HBM3 standards, the memory controller can issue Refresh Management (RFM) commands~\cite{kim2022mithril} to the DRAM for additional mitigative refreshes per tREFI.
We now show how \PRESS{} can leverage these additional mitigations per tREFI to further limit the maximum disturbance.

\begin{figure}[htb]
    \centering
    \includegraphics[width=3.5in]{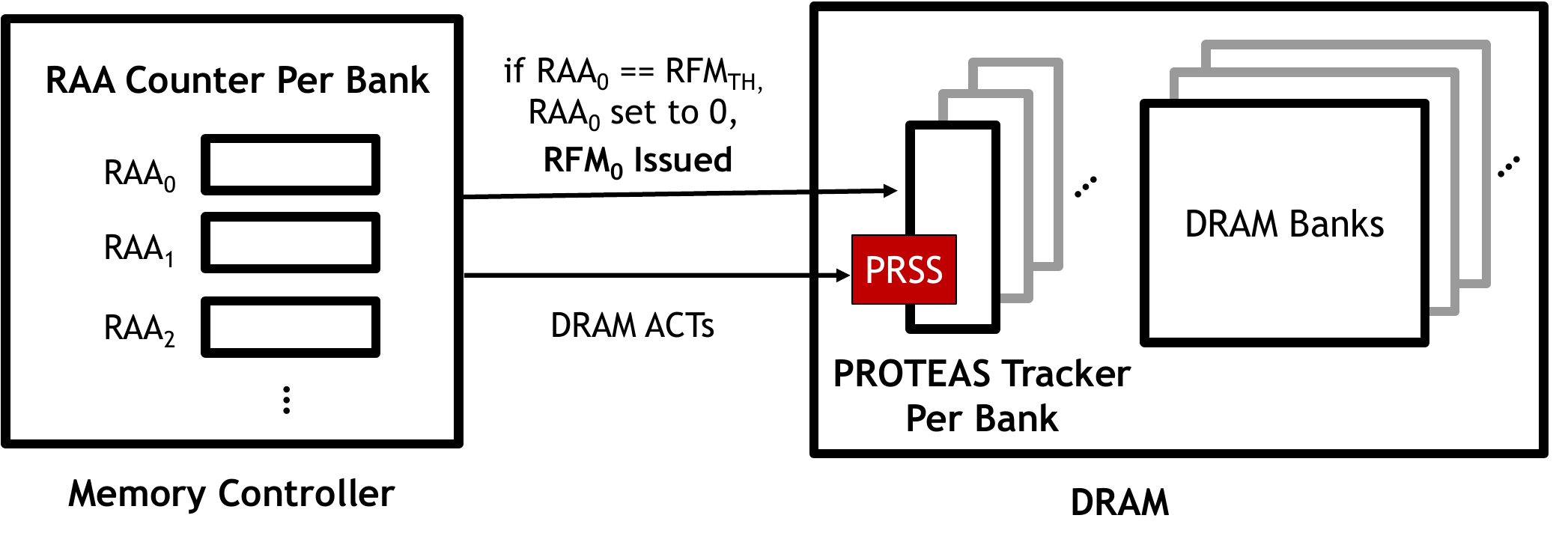}
    \vspace{-0.2 in}
    \caption{Design of PROTEAS with RFM}
    \label{fig:ProteasRFM}
    %\vspace{-0.1 in}
\end{figure}

\cref{fig:ProteasRFM} shows our design of \PRESS{} with RFMs for DDR5 and LPDDR5 (and HBM3).
The memory controller maintains a Rolling Accumulation of ACTs (RAA) counter ~\cite{kim2022mithril}) per bank.
When any RAA counter crosses a threshold, $RFM_{TH}$, the memory controller resets the RAA counter and issues an RFM command for the corresponding DRAM bank.
To ensure $k$ mitigations per tREFI under continuous activations to a DRAM bank, we set the $RFM_{TH}$ to $ACTs\text{-}per\text{-}tREFI/k$, \textit{i.e.},  once every $165/k$ activations. 
On each RFM-based mitigation, as illustrated in \cref{fig:ProbTrackerDesign}, the Mitigation Policy (MFU-based) selects an entry from the tracker and correspondingly issues refreshes to victim rows within the blast-radius (radius of 2 by default).
%The mitigated entry is then invalidated to allow other rows to receive mitigations at subsequent RFMs.

\begin{figure}[!t]
    \centering
    \includegraphics[width=3in]{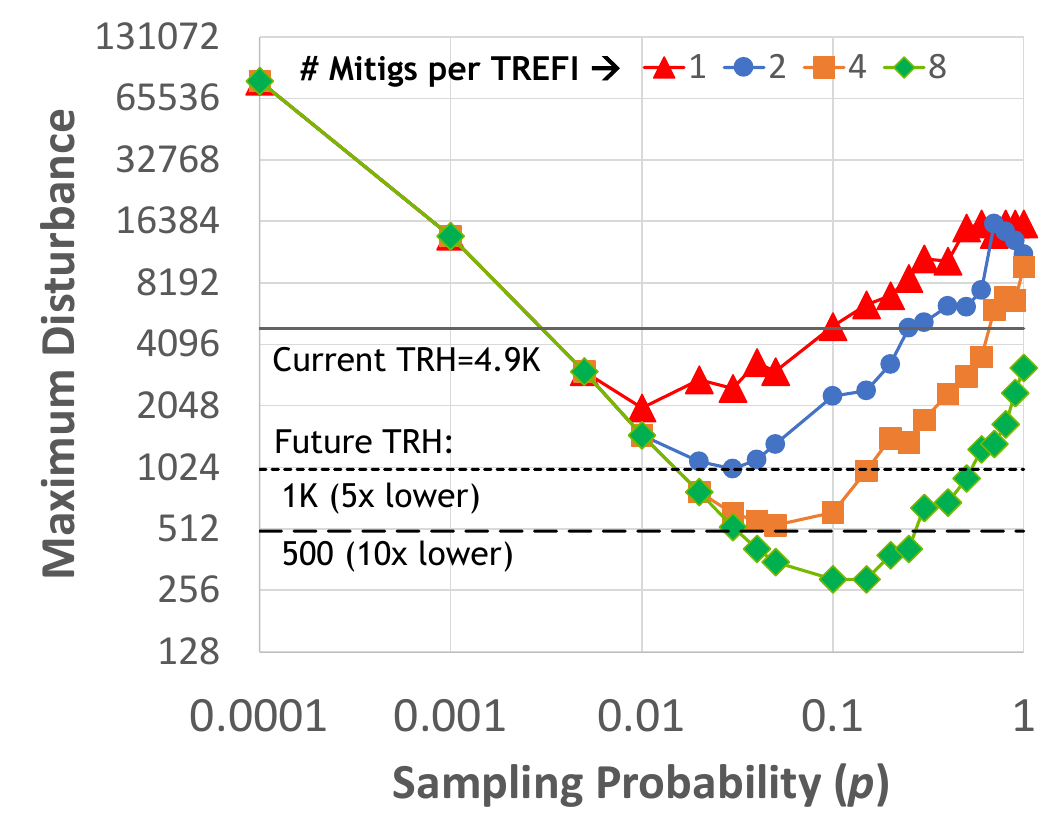}
    \vspace{-0.15 in}
    \caption{Sensitivity of PROTEAS to Number of Mitigations per tREFI. With 4 and 8 mitigations per tREFI, the maximum disturbance decreases below future Rowhammer thresholds of 1K and 500.}
    \label{fig:num_mitig}
    \vspace{-0.1 in}
\end{figure}

\cref{fig:num_mitig} shows how maximum disturbance varies with sampling probability ($p$), as the number of mitigations per tREFI is increased ($RFM_{TH}$ is lowered). 
As the number of mitigations per tREFI is increased from 1 to 2, 4, and 8, the maximum disturbance decreases from 2K to 1K, 533, and 290. 
Additionally, the sampling probability ($p$) at which the minima for the maximum disturbance increases as the frequency of mitigation increases, with it being at $p=0.01$ for the baseline and increasing to 0.03, 0.05, and 0.10 for 2, 4, and 8 mitigations per tREFI. 
This shows that the sampling rate must be proportional to the mitigation rate. If the sampling rate is lower than the mitigation rate, then the proposed policy would be unable to issue mitigations since the tracker can be empty. 
%If the sampling rate is too high, the tracker gets thrashed.
%tracker starts to be fully occupied (and starts to thrash) at a higher sampling probability when the mitigations are being issued more frequently.
Overall, we observe that PROTEAS combined with RFM
%show that increasing the number of mitigations per tREFI to 4 and 8 reduce the maximum disturbance to 533 and 290 respectively. Consequently, such a policy 
is suitable for TRH of 1K and 500.
%in next generation DRAM.
Thus, \PRESS{} is a simple, practical, and scalable solution as Rowhammer thresholds drop to 500 (by 10x compared to 2020 levels).
%\TODO{Can we add 1K and 512 dotted line in the Fig-9}

\begin{comment}
V:  Results
1. Maximum disturbance (1 Mitigation per tREFI) of PRESS + Rand-Repl vs  Graphene-LC, DSAC, ??BIP?? 
    -- Subsection on DSAC - why does it fail?
2. Maximum disturbance (1-N Mitigations per tREFI) of PRESS + Rand-Repl vs  Graphene-LC, DSAC, ??BIP?? 
3. Number of mitigations for achieving same disturbance for  PRESS + Rand-Repl vs  Graphene-LC, DSAC, ??BIP??
3. Comparison of PARA vs PRESS + Rand-Replacement (Same-Mitigation & Same-Disturbance)
4. Performance evaluations of mitigations/tREFI to achieve K-Threshold (PRESS + Rand-Repl & DSAC, Graphene-LC, PARA)
5. Sensitivity of Performance evaluation to Blast-Radius (mitigation cost increases).
\end{comment}

\section{Results}
\subsection{Evaluation Methodology}
We now evaluate \PRESS{} for Rowhammer protection efficacy, performance overheads, and storage overheads.

\begin {table}[!b]
\begin{footnotesize}
\begin{center} 
\vspace{-0.05in}
\caption{Baseline System Configuration}
\vspace{-0.15in}
\begin{tabular}{|c|c|}
\hline
  Out-of-Order Cores           & 4 core, 3GHz, 8-wide fetch, 192 entry ROB   \\
  %Fetch and Retire width & 8         \\ \hline
  Last Level Cache (Shared)    & 4MB, 16-Way, 64B lines \\ \hline
  Memory size, bus speed                  & 16 GB, DDR4, 1.2 GHz (2400 MT/s)  \\
  t$_{RCD}$-t$_{CL}$-t$_{RP}$-t$_{RC}$ & 14.2-14.2-14.2-45 ns\\
  Banks x Ranks x Channels     & 16$\times$1$\times$1 \\
  Rows                & 128K rows, 8KB row buffer\\ \hline
\end{tabular}
\vspace{-0.1in}
\vspace{-0.05 in}
\label{table:system_config}
\end{center}
\end{footnotesize}
\end{table}

\textbf{Rowhammer mitigation efficacy.} We use our Rowhammer simulator to compare \PRESS{} with prior probabilistic mitigations such as DSAC~\cite{DSAC} and PARA~\cite{kim2014flipping}. We collect the maximum disturbance (number of activations before a refresh) across all 500 attack patterns (\cref{sec:analytical_methods}). The reported disturbance is averaged over 100  different seeds provided to the random number generator.
We assume a 16-entry tracker (like TRR) with a baseline policy of deterministic lookup and insertion (100\% sampling).
We configure DSAC to target a TRH of 500. 
We configure the baseline, PARA, and DSAC with similar mitigation frequency as \PRESS{} (1, 2, 4, or 8 mitigations per tREFI). 
For PARA, we achieve a rate of 1, 2, 4, and 8 mitigation per tREFI with mitigation probabilities of 0.6\%, 1.2\%, 2.5\%, and 5\%. 
%\aj{While prior analytical studies~\ref{} report using higher sampling probabilities for PARA at ultra-low TRH, this study takes a quantitative approach for a fair comparison to related work.} 

\textbf{Performance evaluations.} We also model \PRESS{} in Gem5~\cite{lowe2020gem5}, a cycle-accurate simulator in the Syscall Emulation (SE) mode. 
We model a 4-core out-of-order CPU with DDR4 2400MT/s memory, with timings based on Micron datasheets~\cite{micron_ddr4}.
\cref{table:system_config} shows the configuration.
To model the effects of our mitigative action, for 1 mitigation per tREFI configuration, we assume it is issued during tRFC, similar to TRR. 
For additional mitigations per tREFI, we model the overhead similar to RFM, where the bank is busy for a period of $2\times \text{blast-radius} \times tRC$.
We evaluate our design with 17 SPEC2017~\cite{SPEC2017} \textit{rate} workloads and 17 mixed workloads. We fast-forward the workloads by 25 billion instructions to reach regions of interest and simulate for 250 million instructions.

\subsection{Comparing Maximum Disturbance of \PRESS{}}% vs. Probabilistic Defenses}
\begin{figure}[!t]
    \centering
    \includegraphics[width=3.5in]{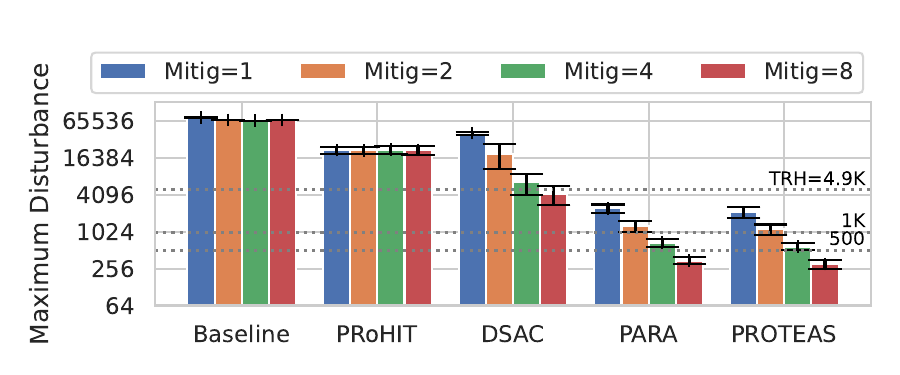}
    \vspace{-0.25in}
    \caption{\textcolor{black}{Maximum Disturbance of \PRESS{} and other probabilistic mitigations. We report the means across \textcolor{black}{100} runs with different seeds and plot error-bars for 95\% confidence intervals.}}
    \vspace{-0.2in}
    \label{fig:max_dist}
\end{figure}

\label{sec:max_dist}
~\cref{fig:max_dist} compares the Max-Disturbance for \PRESS{} with prior probabilistic defenses, \textcolor{black}{PRoHIT~\cite{PROHIT},} DSAC~\cite{DSAC}, and PARA~\cite{kim2014flipping}. We compare  against a baseline deterministic tracker which samples all lookups and uses LFU for  evictions.

At one mitigation per tREFI, \PRESS{} with 1\% sampling achieves a max-disturbance of 2.1K, which is 35X lower than baseline which has a max-disturbance of 74K.
As mitigations per tREFI increase to 2, 4, and 8, the baseline deterministic tracker continues to see a high max-disturbance (65K-67K), whereas \PRESS{} has  max-disturbance that reduces to 1128, 585, and 305 at sampling rates of 3\%, 5\%, and 10\%, respectively.
\PRESS{} achieves 60X to 222X decrease over the baseline because the  deterministic tracker gets easily thrashed despite the additional mitigations per tREFI, and the resident entries are easily evicted between an insertion and potential mitigation, despite the shorter time window with higher mitigation frequency.
However with \PRESS{}, the tracker can get thrashed over a much longer period, and hence frequent mitigations limits the time available to an attacker, thus reducing the maximum disturbance linearly.

\begin{comment}
Mitigs	Baseline	DSAC	PARA	PROTEAS
Mititg/tREFI = 1	73728	40933.5	2461.9	2132.1
Mititg/tREFI = 2	67814	19081.2	1253.9	1128.9
Mititg/tREFI = 4	65536	6483.2	682.5	585.9
Mititg/tREFI = 8	67814	4254.5	348.6	305.2
\end{comment}

In comparison, the prior probabilistic tracker \textcolor{black}{PRoHIT~\cite{PROHIT}, achieves a max-disturbance of 21k with one mitigation per tREFI (10x higher than \PRESS{}). This is because Blacksmith patterns deterministically evict entries from PRoHIT's cold table before they are promoted to the hot table, thus avoiding mitigations for hammered rows.} Similarly, Samsung's DSAC~\cite{DSAC}, achieves a max-disturbance of 41K with one mitigation per tREFI (19x higher than \PRESS{}), which decreases to ~4K at 8 mitigations per tREFI (14x higher than \PRESS{}). 
This is because DSAC adopts stochastic evictions similar to PMSS in \cref{fig:PRSS_PERF_Bathtub}, which reduces thrashing to a less extent compared to PRSS in \PRESS{}.
Moreover, the sampling probability in DSAC varies inversely to the minimum count in the tracker. We observe the $p$ in DSAC dynamically varies between 1 and 0.05 leaving it more vulnerable to thrashing than our PRSS which always operates at the optimal sampling probability.
The highest maximum disturbance for DSAC is achieved for non-uniform attack patterns (where decoy rows are accessed immediately after few iterations of target rows similar to Blacksmith~\cite{jattke2021blacksmith}) which effectively thrash this tracker and severely degrade its security -- these patterns were not evaluated in DSAC.

The maximum disturbance with \PRESS{} is generally lower than PARA at similar mitigation costs. PARA with one mitigation issued per tREFI achieves a max-disturbance of 2.4K (15\% higher than \PRESS{}), and with 8 mitigations per tREFI, this reduces to 350 (14\% higher than \PRESS{}). 
This is because, unlike PARA where the mitigations are fully probabilistic, \PRESS{} samples probabilistically into the tracker and then intelligently selects entries for mitigation based on frequency. This allows \PRESS{} to achieve slightly better resilience at equivalent mitigation costs.
Moreover, unlike PARA which is required to be implemented in the memory-controller, that requires knowledge of neighboring rows in DRAM, \PRESS{} is an in-DRAM solution that can be managed directly by memory vendors.  

\textcolor{black}{Overall, ~\cref{fig:max_dist} shows that with 1 mitigation per tREFI, \PRESS{} reduces max disturbance by 35X versus baseline, \textcolor{black}{by 10X compared to PRoHIT}, by 19X compared to DSAC, and by 15\% compared to PARA. With 8 mitigations/tREFI, these reductions with \PRESS{} are 222X vs baseline, \textcolor{black}{72x vs PRoHIT}, 14X vs DSAC, and 14\% vs PARA.}

%\subsection{Comparison on Mitigations with DSAC and PARA}
%\TODO{PARA and DSAC comparisons.}\aj{let's discuss if we need this or not}

%1. Maximum disturbance (1-N Mitigations per tREFI) of PRESS + Rand-Repl vs  DSAC, PARA (BIP).

%2. Number of mitigations for achieving same disturbance for PRESS, PARA, DSAC, (BIP)

\begin{figure}[!t]
    \centering
    \includegraphics[width=3.5in]{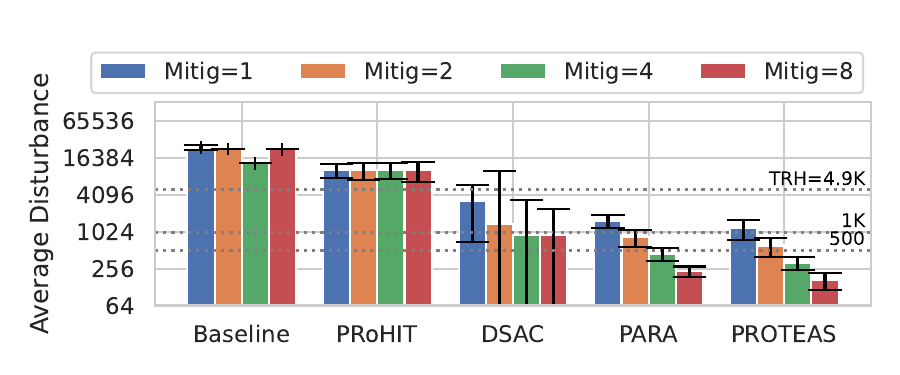}
    \vspace{-0.25in}
    \caption{\textcolor{black}{Average Disturbance of \PRESS{} and other probabilistic mitigations. We report the means across 100 runs with different seeds and plot error-bars for 95\% confidence intervals.}} %With 1 mitigation per tREFI, \PRESS{} reduces average disturbance by 20X versus baseline, \textcolor{black}{by 9X compared to PRoHIT}, by 3X compared to DSAC, and by 30\% compared to PARA. With 8 mitigations /tREFI, these reductions with \PRESS{} are 135X vs baseline, \textcolor{black}{62x vs PRoHIT}, 6X vs DSAC, and 40\% vs PARA.}}
    \vspace{-0.2in}
    \label{fig:avg_dist}
\end{figure}

\subsection{\textcolor{black}{Comparing Average Disturbance of \PRESS{}} } %vs. Probabilistic Defenses}

\textcolor{black}{
We also compare the average disturbance for \PRESS{} to that of \textcolor{black}{PRoHIT~\cite{PROHIT},} DSAC~\cite{DSAC}, and PARA~\cite{kim2014flipping}. Again, we compare these against a baseline deterministic tracker which samples all lookups and uses LFU for  evictions.
Figure~\ref{fig:avg_dist} illustrates the average disturbance received across the different schemes. The figure shows that with 1 mitigation per tREFI, \PRESS{} reduces average disturbance by 20X versus baseline, \textcolor{black}{by 9X compared to PRoHIT}, by 3X compared to DSAC, and by 30\% compared to PARA. With 8 mitigations/tREFI, these reductions with \PRESS{} are 135X vs baseline, \textcolor{black}{62x vs PRoHIT}, 6X vs DSAC, and 40\% vs PARA.
Thus, \PRESS{} reduces both the maximum disturbance and the average disturbance for the hammered rows compared to all prior probabilistic defenses. 
}

\subsection{\textcolor{black}{Sensitivity of \PRESS{} to Tracker Eviction Policy}}

\begin{figure}[!t]
    \centering
    \includegraphics[width=3.5in]{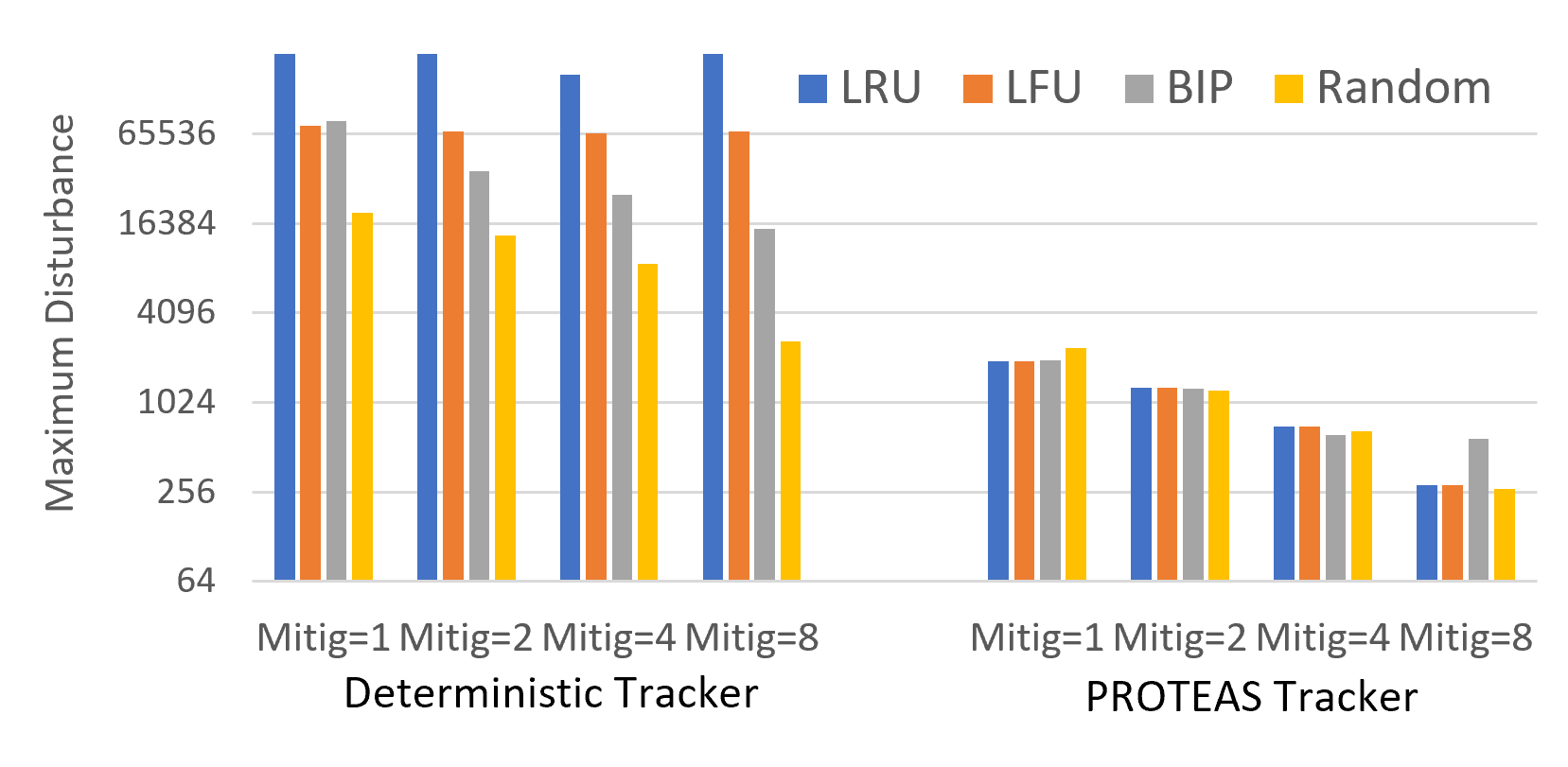}
    \vspace{-0.25in}
    \caption{\textcolor{black}{Sensitivity to Eviction Policy for Baseline and PROTEAS Tracker}}
    \vspace{-0.1in}
    \label{fig:repl_sens}
\end{figure}

\textcolor{black}{To better understand the impact of replacement algorithms on trackers, we study sensitivity to tracker eviction policy in both deterministic and probabilistic trackers. We study LRU, LFU, BIP~\cite{bip}, and random replacement as the eviction policy but maintain a LFU-based mitigation policy. \cref{fig:repl_sens} illustrates the maximum disturbance for the different eviction policies on the baseline deterministic tracker (no sampling based insertion) and a \PRESS{} tracker. For a deterministic tracker, LRU has the highest maximum disturbance (226K) followed by LFU (70K), which remains high even if mitigation frequency is increased because they are easily thrashed. BIP and Random replacement minimize thrashing and reduce maximum disturbance to 15K and 2K with 8 mitigations per tREFI. BIP is thrash-resistant in that it retains a fixed portion of the access stream in the tracker. Random replacement on the other hand retains different portions of the access stream allowing them to receive mitigations over time~\cite{bip}.}

\textcolor{black}{For \PRESS{}, there is limited benefit from intelligent replacement policies since the maximum disturbance across all policies are similar. This is because intelligent replacement policies like LRU/LFU/BIP lose  temporal locality information in the request stream and thus behave like random.}

\begin{figure*}[!tbh]
    \centering
    \includegraphics[width=7.2in]{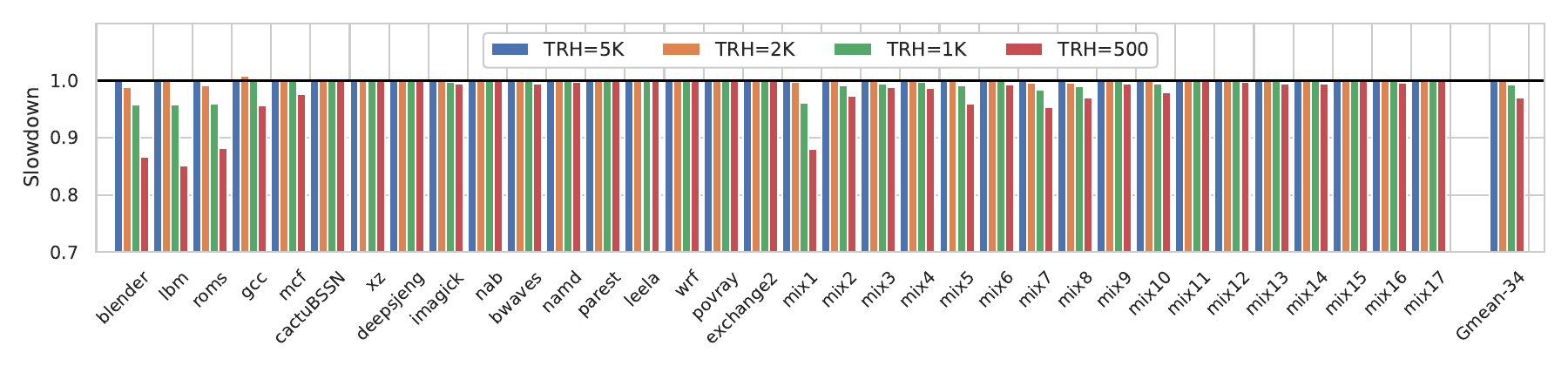}
    \vspace{-0.3 in}
    \caption{Performance Overhead of \PRESS{} (workloads sorted in descending order of rate of DRAM ACTs). \PRESS{} incurs negligible slowdown for TRH of 5K and 2K (1 and 2 mitigations per tREFI), and 0.3\% and 2.9\% slowdown at TRH of 1K and 500 (4 and 8 mitigations per tREFI respectively). 
    }
    \label{fig:perf_proteas}
    \vspace{-0.2 in}
\end{figure*}

\subsection{Performance Overheads of \PRESS{}}
We implement \PRESS{} in Gem5 and measure its performance overheads\footnote{\textcolor{black}{For space reasons we only report performance of \PRESS{}. \PRESS{} and PARA incur similar slowdown. PROTEAS requires 15\% fewer mitigation refreshes to achieve similar Max-Disturbance vs  PARA, so it achieves slightly better performance (but both are within 1\% of each other). Unlike \PRESS{}, PARA is a memory controller solution that cannot be deployed in-DRAM.}} at different Rowhammer thresholds, using 1, 2, 4, and 8 mitigations per tREFI which correspond to TRH values of 5K, 2K, 1K, and 500 respectively. 
As shown in \cref{fig:perf_proteas}, across 17  SPEC-CPU 2017 workloads and 17 mixed workloads (random combinations of SPEC workloads), \PRESS{} incurs negligible slowdown at TRH of 5K and 2K. 
At these operating points, \PRESS{} requires $RFM_{TH}$ of 166 and 83, which results in 1 mitigation every 166 or 83 activations in the worst case. As activations are only a subset of memory accesses, the resulting overheads are insignificant.  

At TRH of 1K and 500, \PRESS{} incurs slowdowns of 0.3\% and 2.9\%. 
The comparatively higher slowdown is because \PRESS{} requires more frequent mitigations at these thresholds (i.e., one mitigation per 42 and 21 activations).
With a default blast-radius of 2, each mitigation incurs 4 activations.
Thus, \PRESS{} at lower thresholds can cause 10\% and 20\% extra DRAM activations, which leads to higher slowdowns.
The highest slowdowns are for workloads like lbm (14.9\%) and blender (13.4\%) at TRH of 500, which have the highest DRAM activation rates of more than 20 per thousand instructions (PKI).
Workloads such as cactuBSSN and those after it in the sorted list in \cref{fig:perf_proteas} have a DRAM activation rate of less than 1 PKI and thus incur negligible slowdown even at lower thresholds.

\begin{comment}
% Column1	PRESS	PARA

% TRH=5000	1	    0.981
% TRH=1000	0.999	??
% TRH=1000	0.993	0.942
% TRH=500	0.97	0.828

\begin{figure}[!htb]
    \centering
    \includegraphics[width=3in]{Graphs/Perf.TRH.pdf}
    \caption{Performance of \PRESS{} and PARA, at different Rowhammer Thresholds (TRH). At TRH=2000, \PRESS{} has negligible slowdown while PARA has 2\% slowdown, but as TRH reduces to 1000 and 500, \PRESS{} has a slowdown of 0.7\% and 3\%, whereas PARA has a slowdown of 6\% and 17\%.}    
    \label{fig:perf_thresholds}
\end{figure}
\end{comment}
\begin{figure}[!htb]
    \centering
    \includegraphics[width=3.4in]{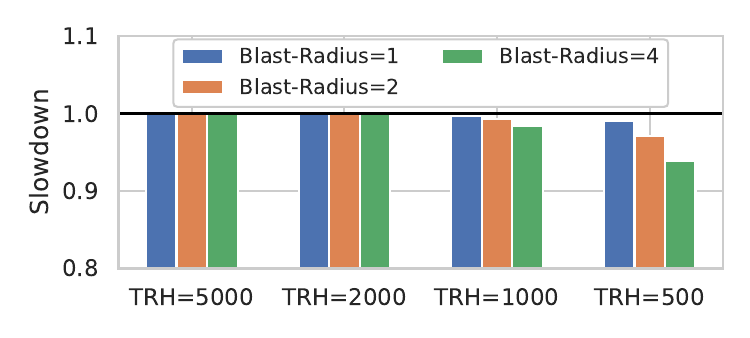}
    \vspace{-0.1in}
    \caption{Performance of \PRESS{} as Blast Radius (BR). \PRESS{} incurs negligible slowdown at TRH of 5K and 2K. At TRH of 1K, as Blast-Radius increases from 1 to 2 to 4, \PRESS{} slowdown increases from 0.3\%, 0.7\% and 1.5\%, while at TRH=500, the slowdown increases from 1\% to 3\% to 6\%.}
    \vspace{-0.1in}
    \label{fig:perf_br}
\end{figure}
\subsection{Sensitivity of Performance to Blast Radius}

\cref{fig:perf_br} shows the performance-sensitivity of \PRESS{} to the blast-radius of the mitigation. 
Our default defense refreshes victim rows within a blast radius of 2 from the aggressor (refreshes to 2 rows above and 2 rows below) to ensure protection against Half-Double~\cite{HalfDouble} attacks which flip bits in distance-2 victims.
At TRH of 5K and 2K, we observe negligible slowdown even for blast radius up to 4. 
At TRH of 1K, as blast-radius of mitigation increases from 1 to 2 to 4, \PRESS{} slowdown increases from 0.3\%, 0.7\%, and 1.5\%, whereas at TRH of 500, the slowdown increases from 1\% to 3\% to 6\%.
As mitigation blast radius increases from 1 to 4, the number of neighbor rows refreshed per mitigation increases from 2 to 8. 
This increases the slowdown at lower thresholds.

\begin{comment}
        TRH  Blast-Radius=1  Blast-Radius=2  Blast-Radius=4
0  TRH=5000           1.000           1.000           1.000
1  TRH=2000           0.999           0.999           0.999
2  TRH=1000           0.997           0.993           0.984
3   TRH=500           0.990           0.971           0.939
\end{comment}

\begin{comment}
% Column1	PRESS	PARA
% BR=1	0.997	0.977
% BR=2	0.993	0.942
% BR=4	0.985	0.886

\begin{figure}[!htb]
    \centering
    \includegraphics[width=3in]{Graphs/Perf.BR.pdf}
    \caption{Performance of \PRESS{} and PARA, at different Blast Radius (BR) for TRH=1000. As mitigation blast radius increases from  1 to 2 and 4, slowdown for \PRESS{} increases from 0.3\% to 0.7\% and 1.5\%, whereas  slowdown for PARA increases from 2\% to 6\% and 11\%.}    
    \label{fig:perf_br}
\end{figure}
\end{comment}

%\subsection{Comparisons with Deterministic Solutions}
%\TODO{Quantify the storage overhead of PROTEAS. Perhaps qualitative comparison (table) with other deterministic trackers can be in related work?.}

\subsection{Storage Overheads of \PRESS{}}
\label{sec:design_overhead}
\PRESS{} requires 16 counters (each requiring 5-bytes: 21  bit counter and 17-bit rowID) per DRAM-bank (\cref{table:storage}). 
In DDR4, across 16 banks, this incurs a storage overhead of 1.3 KB per rank, and in DDR5 (32 banks) requires 2.6 KB of counters  per rank in-DRAM. 
In the memory controller, RFM requires one 8-bit counter per DRAM bank, incurring 32B of storage within the memory controller for 32 banks per rank in DDR5. 
These  overheads are similar to TRR and RFM.

\begin{table}[!htb]
  \centering
%  \vspace{-0.1in}
  \caption{Storage Overheads of \PRESS{}}
  \vspace{-0.05in}
  \begin{footnotesize}
  \label{table:storage}
  \begin{tabular}{cccc}
    \hline
    \multirow{2}{*}{\textbf{Structure}} & \multirow{2}{*}{\textbf{Per-Bank}} & \multicolumn{2}{c}{\textbf{Per-Rank}}  \\ 
    & &  \textbf{DDR4} & \textbf{DDR5} \\ \hline
    \rule{0pt}{0.8\normalbaselineskip}
    Tracker in DRAM           & 16 x 40-bits & 1.3KB  & 2.6KB \\ 
    RFM Counters in MC      & 1 x 8-bits & - & 32B  \\ \hline 
  \end{tabular}
  \end{footnotesize}
\end{table}

\PRESS{} also requires extra logic for two PRNGs per DRAM bank (for PRSS and random replacement), requiring a few thousand gates~\cite{hwprng} in-DRAM per bank. We believe this is practical especially since recent work by Hynix~\cite{HynixRH} showcased a DRAM chip fabricated with a PRNG per bank. %\TODO{What did Hynix use the PRNG for? [GS: For probabilistic sampling of insertions into the tracker.]}\aj{}

\subsection{Energy Overheads of \PRESS{}}
\PRESS{} relies on mitigative refreshes to victim rows to prevent Rowhammer, which can lead to an increase in DRAM energy consumption. 
On one hand, the additional mitigations require extra activation and precharge operations internally within the DRAM, which increase dynamic energy consumption.
On the other hand, extra mitigations may also cause the overall execution time to increase, which causes an increase in static DRAM energy (due to normal refreshes and leakage power).
\cref{table:energy} shows the average energy overheads for \PRESS{} across 34 workloads reported from Gem5, compared to a baseline without extra mitigations.
On average, \PRESS{} incurs negligible energy overheads at TRH of 5K, whereas at TRH of 1K, this overhead goes up to 0.5\%, and 2.4\% at TRH of 500. 
While the dynamic energy increases by 4.2\% at TRH of 500, the static energy, which makes up the majority of the DRAM energy for several workloads, increases by only 2.3\%, making the overall increase in energy of 2.4\%. 

\begin{comment}
	Static Energy	Dynamic Energy	Total Energy
R1	0.0%	0.0%	0.0%
R4	0.6%	0.5%	0.5%
R8	2.3%	4.2%	2.4%
\end{comment}
\begin{table}[!htb]
  \centering
  \vspace{-0.1in}
  \caption{DRAM Energy Overheads of \PRESS{} (blast-radius $=$ 2)}
  %\vspace{0.1in}
  \begin{footnotesize}
  \label{table:energy}
  \begin{tabular}{cccc}
    \hline
    \textbf{\PRESS{} Config} & \textbf{Static Energy} & \textbf{Dynamic Energy} & \textbf{Total Energy}  \\ \hline
    \rule{0pt}{0.8\normalbaselineskip}
    TRH $=$ 5K           & 0.0\% & 0.0\%  & 0.0\% \\ 
    TRH $=$ 1K           & 0.6\% & 0.5\%  & 0.5\% \\ 
    TRH $=$ 500            & 2.3\% & 4.2\%  & 2.4\% \\ \hline 
  \end{tabular}
   \vspace{-0.1in}
  \end{footnotesize}
\end{table}

\section{Related Work}

% SK: Following paragraph is fluff - not necessary
% Our Rowhammer defense focuses on detection of aggressor-rows using counter-based trackers and issuing mitigative refreshes to neighboring victims.  Thus, we now discuss prior art that are closely related to our work. 

\subsection{Deterministic Aggressor-Row Trackers}
% Two extremities (Graphene, Mithril) and (Hydra).  
Prior Rowhammer detection mechanisms use resource intensive trackers to identify frequently accessed DRAM rows deterministically. These structures can either be stored on-chip in SRAM or off-chip within the DRAM itself. 
\cref{table:CompareRelatedWork} compares the storage overheads of these trackers with \PRESS{}.

At one end of the spectrum, Graphene~\cite{park2020graphene} uses the Misra-Gries algorithm for tracking, which provides a lower bound on the number of counters required for guaranteed detection of rows that may be activated beyond the Rowhammer threshold.
Mithril~\cite{kim2022mithril} and PROTRR~\cite{ProTRR} leverage a similar algorithm and store such a tracker within the logic portion of the DRAM.
%Unfortunately, the storage overheads for such solutions scales with the number of DRAM banks, and inversely with TRH.
However, at TRH of 500, and with the doubling of number of banks in DDR5, such structures require a SRAM storage of 640KB per rank, which is too high to be stored in SRAM on-chip~\cite{park2020graphene} or in the logic portion of the DRAM~\cite{kim2022mithril}. 

At the other end of the spectrum, Hydra~\cite{qureshi2022hydra} (and CRA~\cite{kim2014architectural}) stores one counter per row in DRAM, with additional filters or caches in on-chip SRAM to limit extra accesses to DRAM. 
While such trackers require reserving 2.3MB of storage in DRAM, which is not significant, the SRAM-based performance optimizations are access-pattern dependent, which can lead to a worst-case slowdown of up to 70\% with pathological workloads.
Storing such counters in the DRAM array~\cite{bennett2021panopticon,HynixRH} to avoid these performance issues requires a significant redesign of the DRAM arrays, which may affect DRAM access times, and is less desirable. 

Other prior trackers typically incur different storage and performance overheads.
TWiCe~\cite{lee2019twice} maintains a table of counters to track activations per row, and prunes the entries that are unlikely to reach the Rowhammer threshold to save space. 
CAT~\cite{CATT} maintains a dynamic tree of counters which allocates more counters to hot rows, to enable a storage-efficient tracker.
The storage overheads of such solutions scale as the number of attacked rows increase. 
At low thresholds of 500, such solutions incur higher storage overheads than Hydra or CRA, making them impractical.

D-CBF~\cite{yauglikcci2021blockhammer} uses a counting bloom-filter to identify potential aggressor rows that cross a certain threshold of activations. Unfortunately, as this is a blocklisting-based tracker, rows once inserted into the filter will continue to be flagged as aggressors until the end of the refresh period (64ms) when the tracker can be reset, making it incur high mitigation costs. Additionally, this incurs a high SRAM storage overhead of 1.5MB per rank. 

In contrast, \PRESS{} incurs negligible storage costs of less than 3KB SRAM stored within the DRAM logic area (equivalent to currently deployed TRR) and has low mitigation costs, while providing significantly better security than TRR by enabling effective thrash protection. 
%\PRESS{} even provides considerably strong security compared to recent resource-constrained probabilstic trackers like Samsung's DSAC, which we \PRESS{} provides significant   

\ignore{
Existing row hammer detection mechanisms maintain counter-based data structures to track frequently access DRAM rows.  These data structures can either be stored on-chip in SRAM or off-chip within the DRAM array itself.  In-DRAM array data structures require changes to the existing DRAM structure and also incur additional bandwidth to retrieve and update the counter values.  Comparatively, on-chip counter-based structures are either placed directly within the memory-controller or in the logic portion of the DRAM. These structures tend to be small and have generally been suitable for large row hammer thresholds ($TRH>16K$). Scaling these data structures for ultra low row hammer thresholds (i.e., $TRH <=500$) require impractical data structure sizes. For ultra low row hammer thresholds ($TRH=500$), \cref{table:storage} illustrates the storage overheads of recent proposals.  Compared to prior art, PROTEAS requires nearly $20-2000x$ lower storage overhead. In fact, PROTEAS incurs a negligible 1.3KB overhead on DDR4 and 2.6KB overhead on DDR5 systems.
}

%\vspace{-0.2in}
\begin{table}[!htb]
%\vspace{0.1in}
%\renewcommand{\arraystretch}{1.2}
  \caption{Storage Overheads of Trackers per 16GB DRAM Rank at TRH of 500 (all storage is in SRAM unless specified otherwise)}
  
  \centering
  \begin{footnotesize}
  \label{table:CompareRelatedWork}
  \begin{tabular}{lcc}
    \hline
    \textbf{Name} & \textbf{DDR-4} (16 banks/rank)&\textbf{DDR-5} (32 banks/rank)  \\ \hline
    \multirow{2}{*}{Hydra} & 26 KB (SRAM) & 52 KB (SRAM) \\
                           & 2.3MB (DRAM) & 2.3MB (DRAM) \\
    TWiCe& 2.3 MB& 4.6 MB \\
    CAT& 1.5 MB & 3.0 MB \\
    D-CBF& 768 KB & 1.5 MB \\
    Graphene& 340 KB& 640 KB \\
    
    \textbf{PROTEAS}& \textbf{1.3 KB} (in DRAM) & \textbf{2.6 KB} (in DRAM)\\ \hline
  \end{tabular}
  \end{footnotesize}
 \vspace{-0.1in}
\end{table}

\subsection{Alternative Rowhammer Mitigation Mechanisms}
\PRESS{} uses a mitigative action of refreshing the neighboring victim rows~\cite{hassan2021uncovering, 
 park2020graphene, kim2022mithril,DSAC, qureshi2022hydra}.
Alternatively, REGA~\cite{REGA_SP23} proposed a mitigation of refreshing additional rows in parallel to DRAM activations, at no slowdown, by modifying the DRAM sense amplifiers. 
This avoids slowdown but has the cost of significant additional refreshes and increased DRAM energy consumption.
\PRESS{} can be combined with REGA, to invoke such refreshes selectively only for the neighborhood of aggressor rows, thus significantly lowering the DRAM energy overheads, while enjoying the performance benefits of simultaneous refresh and activation.

Prior works have also proposed alternative mitigation that penalizes the aggressors.
%For instance, 
%Existing row hammer mitigation mechanisms take different approaches to address row hammer.  The most common mechanism, including the one studied in our approach, employs victim refresh where rows adjacent to aggressor rows are issued mitigations to restore their charge. Rather than targeting victim rows, some 
For instance, recent schemes like RRS~\cite{saileshwar2022RRS} and SRS~\cite{SRS} move aggressor rows within memory by periodically swapping them with another random row in memory. 
AQUA~\cite{AQUA} similarly moves an aggressor row to a quarantine region once it has been activated beyond a certain threshold.
As such mitigative actions are invoked from the memory-controller, they cannot be directly applied with \PRESS{}, which is an in-DRAM defense.
On other hand, in-DRAM mitigations like SHADOW~\cite{ShadowHPCA23} or CROW~\cite{crow} which similarly relocate aggressor or victim rows in-DRAM within the sub-array at low cost can be combined with \PRESS{}. 

Alternatively, memory-controller-based solutions like Blockhammer~\cite{yauglikcci2021blockhammer} limit the access rate to potential aggressor rows. Such mechanisms incur significant worst-case performance overheads and as such cannot be combined with \PRESS{} given that it is implemented within DRAM.

\subsection{Cryptographic Detection of Rowhammer Bit-Flips}
Recent works propose storing cryptographic signatures like Message Authentication Codes (MACs) for data in DRAM, and using them to verify that data is free from tampering on DRAM reads.
%before consuming such data.
SafeGuard~\cite{ali2022safeguard} and CSI-RH~\cite{csi} store MACs for each 64 byte data in DRAM, 
%in the ECC-chips in ECC-DIMMs
whereas PT-Guard~\cite{DSN23_PTGuard} only stores MACs for OS page-table data.
%to work with even non-ECC memories. 
Since these solutions detect a Rowhammer attack when bits flip, uncorrectable multi-bit flips can lead to data loss.
On the other hand, because tracker-based solutions like \PRESS{} adopt an orthogonal approach of tracking aggressors and issuing mitigations to prevent a Rowhammer bit-flip, such data loss is prevented.

\vspace{0.1in}
\section{Conclusion}

%Academia and industry have invested significant research effort in addressing the row hammer problem. Memory vendors desire solutions with low storage overhead, low performance overheads and strong security. As such, 

In current DRAM modules, DRAM vendors have deployed Target Row Refresh (TRR) which uses a small in-DRAM tracking structure with tens of counters to detect row hammer attacks. Unfortunately, thrashing-based attack patterns 
%using uniform and non-uniform access patterns 
have rendered TRR vulnerable. 
This paper proposes  \emph{\underline{PRO}babilistic \underline{T}rack\underline{E}r m\underline{A}nagement policie\underline{S} (\PRESS{})} to make in-DRAM trackers thrash resistant. \PRESS{} employs Probabilistic Request Stream Sampling (PRSS) and random replacement to prevent thrashing and introduce diversity in entries held within the tracker. \PRESS{} significantly reduces the maximum disturbance (i.e., number of activations before a mitigation) by 35X compared to a deterministic tracker baseline at current TRH of 4.9K, and by 220X at thresholds of 500 when co-designed with RFM. 
%We show that \PRESS{} alone is suitable for conventional DRAMs with 4.9K row hammer threshold. To further reduce maximum disturbance for DRAMs with ultra low row hammer threshold ($TRH<=500$), we co-design ~\PRESS{} with Refresh Management (RFM) on DDR5 systems. We show that increasing the number of mitigations per TREFI using RFMs are best suited for PROTEAS managed tracker. We show that issuing 8 mitigations per TREFI reduces maximum disturbance by $230X$ compared to baseline. 
PROTEAS requires only 2.6KB storage overhead on DDR5 systems while incurring only 0.3\% and 3\% slowdown for TRH of 1K and 500, respectively.

%\appendix
\begin{appendices}
\section{\textcolor{black}{Analytically Deriving Optimal Sampling Rates}}

\ignore{
\textcolor{black}{
The goal of \PRESS{} is to minimize tracker thrashing. So we operate \PRESS{} at sampling rates that minimizes thrashing.
}
}

\textcolor{black}{
All entries inserted into the tracker exit the tracker either when they are evicted on a capacity/conflict miss OR when the entry is mitigated. \PRESS{} desires to operate at a sampling probability where thrashing is minimized such that entries inserted exit from the tracker predominantly on mitigations. At this point, the  tracker insertion rate should be at least equal to the tracker mitigation rates, as inserting at a lower rate than the mitigation rate means that the tracker is not fully utilized. The insertion rate should be no more than the mitgation rate, as that would overflow the tracker and cause thrashing.
}

\textcolor{black}{
For sequence of $A$ activations to a tracker with sampling rate $S$ and a given $MissRate$, the total insertions ($I_A$) is:
}
\begin{equation}
 \textcolor{black}{
 I_A = A \times MissRate \times S
}
\end{equation}

\textcolor{black}{
The total mitigations ($M_A$) for $A$ activations assuming  mitigation rate $M$ (number of mitigations per activation) is:
}

\begin{equation}
\textcolor{black}{
      M_A = A \times M 
}
\end{equation}

\textcolor{black}{
To avoid thrashing, ~\PRESS{} desires $I_A$ equal $M_A$. By equating (1) and (2), the sampling rate $S$ is:
}

\begin{equation}
\textcolor{black}{
S = M / MissRate 
}
\end{equation}

\textcolor{black}{
Without loss of generality, assuming a $MissRate = 0.5$; the optimal analytical sampling rate of \PRESS{} is = $M \times 2$. ~\cref{table:analytical} 
%compares the analytical rate to the best performing sampling rate observed empirically.  We see 
shows that the empirical sampling rate closely matches the analytical sampling rate.
}

\begin{table}[htb]
  \centering
  \begin{footnotesize}
    %\vspace{-0.1in}
  \caption{\textcolor{black}{Sampling Rates Analytical vs Empirical}}
 \vspace{-0.1in}
  \label{table:analytical}
  \begin{tabular}{ccc}
    \hline
    \multirow{ 2}{*}{\textcolor{black}{Mitigation Rate}} & \textcolor{black}{Sampling Rate} & \textcolor{black}{Sampling Rate}\\ 
                                        & \textcolor{black}{(analytical)} & \textcolor{black}{(empirical)} \\   \hline
    \textcolor{black}{1 Mitigation per 166 ACTs} & \textcolor{black}{1.2} & \textcolor{black}{1}\\
    \textcolor{black}{2 Mitigation per 166 ACTs} & \textcolor{black}{2.4} & \textcolor{black}{3}\\
    \textcolor{black}{4 Mitigation per 166 ACTs} & \textcolor{black}{4.8} & \textcolor{black}{5}\\
    \textcolor{black}{8 Mitigation per 166 ACTs} & \textcolor{black}{9.6} & \textcolor{black}{10}\\
  \end{tabular}
  \vspace{-0.2in}
  \end{footnotesize}
\end{table}

\end{appendices}

%%%%%%% -- PAPER CONTENT ENDS -- %%%%%%%%

%%%%%%%%% -- BIB STYLE AND FILE -- %%%%%%%%
\newpage
\bibliographystyle{IEEEtranS}
\bibliography{refs}
%%%%%%%%%%%%%%%%%%%%%%%%%%%%%%%%%%%%

\end{document}